%% file: paper.tex
\documentclass[a4paper,11pt]{article}

\usepackage{acronym}
\usepackage[american]{babel}
\usepackage[utf8x]{inputenc}
\usepackage{xspace}
\usepackage{amssymb}
\usepackage{amsmath}
\usepackage{authblk}
\usepackage[top=1in,left=1in,bottom=1in,right=1in]{geometry}
\usepackage{graphicx}
\usepackage{caption}
\usepackage{subcaption}
\usepackage{dsfont}
\usepackage{color}
\usepackage{multirow}
\usepackage[pdfborder={0 0 0}]{hyperref} 
\def\todo#1{}

\input{acronyms}
\input{commands}
\input{cps-commands}

\title{\textbf{Engineering Art Galleries}}

\author[1]{Pedro J.\ de Rezende}
\author[1]{Cid C.\ de Souza}
\author[2]{Stephan Friedrichs}
\author[3]{Michael Hemmer}
\author[3]{Alexander Kr\"oller}
\author[1]{Davi C.\ Tozoni}

\affil[1]{University~of~Campinas,~Institute~of~Computing,~Campinas,~Brazil\\
	E-Mail:~\texttt{\{rezende,cid\}@ic.unicamp.br},~\texttt{davi.tozoni@gmail.com}}
\affil[2]{Max~Planck~Institute~for~Informatics,~Saarbr\"ucken,~Germany\\
	E-Mail:~\texttt{sfriedri@mpi-inf.mpg.de}.}
\affil[3]{TU~Braunschweig,~IBR,~Algorithms~Group,~Braunschweig,~Germany\\
	E-Mail:~\texttt{mhsaar@gmail.com},~\texttt{kroeller@perror.de}}

\date{}

\begin{document}

\maketitle

\begin{abstract}
  The \acl{AGP} is one of the most well-known problems in
  \acl{CG}, with a rich history in the study of
  algorithms, complexity, and variants. Recently there has been a
  surge in experimental work on the problem. In this survey, we
  describe this work, show the chronology of developments, and compare
  current algorithms, including two unpublished versions, in an exhaustive
  experiment. Furthermore, we show
  what core algorithmic ingredients have led to recent successes.

\end{abstract}

\input{intro}

\acresetall
\input{problem}

\input{time}

\input{experiments}

\input{speedups}

\input{open}

\acresetall
\input{conclusion}

\section*{Acknowledgments}

Many people have contributed to the developments described in this
paper. In particular, the authors would like to thank Tobias
Baumgartner, Marcelo C.~Couto, S{\'a}ndor P.~Fekete, Winfried
Hellmann, Mahdi Moeini, Eli Packer,
and Christiane Schmidt.

Stephan Friedrichs was affiliated with TU Braunschweig, IBR during most of the research.

This work was partially supported by the Deutsche
Forschungsgemeinschaft~(DFG) under contract number KR~3133/1-1
(Kunst!), by \cpsackFAPESP, \cpsackCNPq, and \cpsackFAEPEX. Google Inc.\ supported the development of the \cgalcite
visibility package through the 2013 Google Summer of Code.

\bibliographystyle{abbrv}
\bibliography{refs,cps-refs}

\end{document}

%% file: acronyms.tex
\acrodef{AGP}{Art Gallery Problem}
\newcommand{\agp}{\ac{AGP}\xspace}

\acrodef{AVP}{Atomic Visibility Polygon}
\newcommand{\avp}{\ac{AVP}\xspace}
\newcommand{\avps}{\acp{AVP}\xspace}

\acrodef{CGAL}{Computational Geometry Algorithms Library}
\newcommand{\cgal}{\ac{CGAL}\xspace}
\newcommand{\cgalcite}{\acl{CGAL}~\cite{cgal} (\acs{CGAL}\acused{CGAL})\xspace}

\acrodef{CG}{Computational Geometry}
\newcommand{\cg}{\ac{CG}\xspace}

\acrodef{CP}{Chwa Points}
\newcommand{\cp}{\ac{CP}\xspace}

\acrodef{CPLEX}{IBM ILOG CPLEX Optimization Studio}
\newcommand{\cplex}{\ac{CPLEX}\xspace}
\newcommand{\cplexcite}{\acl{CPLEX}~\cite{cplex} (\acs{CPLEX}\acused{CPLEX})\xspace}

\acrodef{CV}{Convex Vertices}
\newcommand{\cv}{\ac{CV}\xspace}

\acrodef{ILP}{Integer Linear Program}
\newcommand{\ilp}{\ac{ILP}\xspace}

\acrodef{LP}{Linear Program}
\newcommand{\lp}{\ac{LP}\xspace}

\acrodef{SC}{Set Cover}

\acrodef{SCP}{Set Covering Problem}
\newcommand{\scp}{\ac{SCP}\xspace}

%% file: commands.tex
\def\etal{et~al.}

\newcommand{\wo}{\setminus}

\newcommand{\ie}{i.e.\xspace}

\def\dsR{\mathds{R}}

\def\vissym{\mathcal{V}}
\newcommand{\vis}[1]{\operatorname{\vissym}\left(#1\right)}

\def\AGP{\operatorname{AGP}}
\def\AGPFrac{\operatorname{AGP}_{\mathds{R}}}

\newcommand{\ignore}[1]{}

\def\visarr{\mathcal{A}}

\def\bigO{O}

\newcommand{\algversion}[1]{{\tt #1}\xspace}
\newcommand{\bsx}{\algversion{BS-2010}}
\newcommand{\bsxii}{\algversion{BS-2012}}
\newcommand{\bscur}{\algversion{BS-2013}}

\newcommand{\cix}{\algversion{C-2009}}
\newcommand{\cxiiisea}{\algversion{C-2013.1}}
\newcommand{\cxiiifull}{\algversion{C-2013.2}}
\newcommand{\ccur}{\algversion{C+BS-2013}}

%% file: cps-commands.tex
\newcommand{\cpsackCNPq}{Conselho Nacional de Desenvolvimento Cient\'\i fico e Tecnol\'ogico
({\sc CNP}q, grants \#311140/2014-9, \#477692/2012-5 and \#302804/2010-2)} 
\newcommand{\cpsackFAPESP}{Fun\-da\-\c c\~ao de Amparo \`a Pesquisa do Estado de S\~ao Paulo ({\sc Fapesp}, \#2007/52015-0, \#2012/18384-7)} 
\newcommand{\cpsackFAEPEX}{F{\sc aepex}/U{\sc nicamp}}

\definecolor{darkgreen}{rgb}{0,0.5,0}
\newcommand{\dctC}[1]{#1}

\newcommand{\cidC}[1]{#1}

%% file: intro.tex
\section{Introduction}
\label{sec:intro}

\begin{figure}
  \centering
  \begin{subfigure}[b]{0.45\textwidth}
    \includegraphics[width=\textwidth]{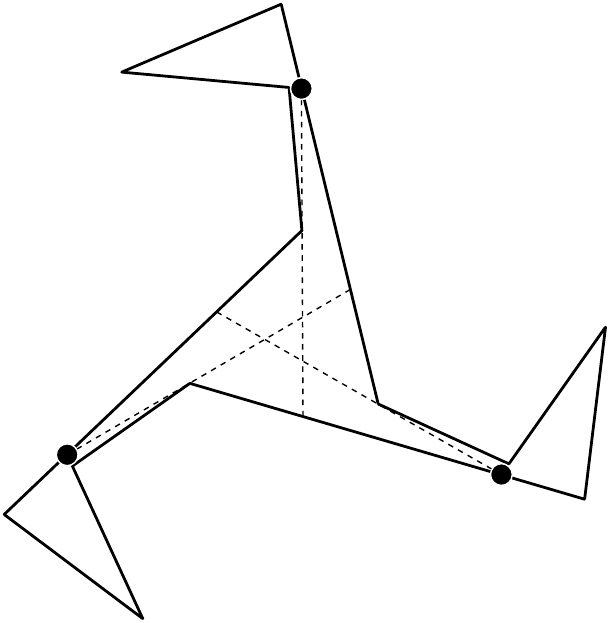}
    \caption{Three guards suffice to cover the walls, but not the interior.}
    \label{fig:wall-vs-interior}
  \end{subfigure}
  ~
  \begin{subfigure}[b]{0.45\textwidth}
    \includegraphics[width=\textwidth]{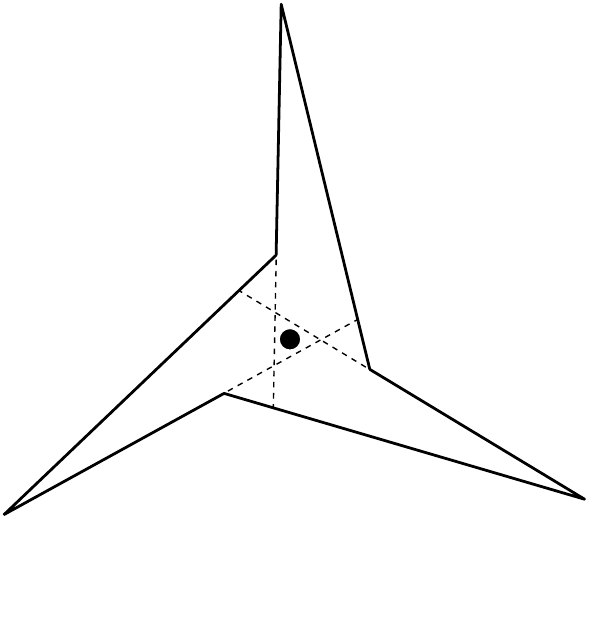}
    \caption{One point guard covers the interior, but a vertex guard cannot}
    \label{fig:godfried}
  \end{subfigure}
  \caption{Edge cover and vertex guard variants have better and worse solutions
    than the classic \agp, respectively.}
\end{figure}

The \agp is one of the classic problems in
\cg. Originally it was posed forty years ago, as recalled by
Ross Honsberger~\cite[p.104]{h-mg2-76}:
\begin{quote}
``At a conference in Stanford in August, 1973, Victor Klee asked the gifted young
Czech mathematician V{\'a}clav Chv{\'a}tal (University of Montreal) whether he had
considered a certain problem of guarding the paintings in an art gallery. The way
the rooms in museums and galleries snake around with all kinds of alcoves and
corners, it is not an easy job to keep an eye on every bit of wall space. The
question is to determine the minimum number of guards that are necessary to
survey the entire building.''
\end{quote}
It should be noted that a slightly different definition is used today,
where not only the walls of the gallery have to be guarded, but also
the interior (this is indeed a different
problem, see Figure~\ref{fig:wall-vs-interior}). \agp has received enormous attention from the
\cg community, and today no CG textbook is complete
without a treatment of it. We give an overview on the most
relevant developments in Section~\ref{sec:problem}, after introducing
the problem more formally.

Besides theoretical interest, there are practical problems that
turn out to be \agp. Some are of these are straightforward, such as
guarding a shop with security cameras, or illuminating an environment
with few lights. For another example, consider a commercial service
providing indoors laser scanning: Given an architectural drawing of an
environment, say, a factory building, a high-resolution scan needs to
be obtained. For that matter, the company brings in a scanner, places
it on a few carefully chosen positions, and scans the building. As
scanning takes quite a while, often in the range of several hours per
position, the company needs to keep the number of scans as low as
possible to stay competitive --- this is exactly minimizing the number
of guards (scan positions) that still survey (scan) the whole
environment.

In this paper, we provide a thorough survey on experimental work in
this area, i.e., algorithms that compute optimal or good solutions for
\agp, including some problem variants. We only consider algorithms
that have been implemented, and that underwent an experimental
evaluation. During the past seven years, there have been tremendous
improvements, from being able to solve instances with tens of vertices
with simplification assumptions, to algorithm implementations that find optimal
solutions for instances with several thousands of vertices, in
reasonable time on standard PCs. We avoid quoting experimental results
from the literature, which are difficult to compare to each other due
to differences in benchmark instances, machines used, time limits, and
reported statistics. Instead, we conducted a massive unified experiment with
900 problem instances with up to 5000 vertices, comparing six
different implementations that were available to us. This allows us to
pinpoint benefits and drawbacks of each implementation, and to exactly
identify where the current barrier in problem complexity lies. Given
that all benchmarks are made available, this allows future work to
compare against the current state. Furthermore, for this paper, the
two leading implementations were improved in a joint work between
their respective authors, using what is better in each. The resulting
implementation significantly outperforms any previous work, and
constitutes the current frontier in solving \agp.

\bigskip

The remainder of this paper is organized as follows. In the next
section, we formalize the problem and describe related
work. In Section~\ref{sec:time}, we turn our attention to the sequence
of experimental results that have been presented in the past few
years, with an emphasis on the chronology of developments. This is followed by an experimental cross-comparison of
these algorithms in Section~\ref{sec:experimental_evaluation}, showing speedups over
time, and the current frontier. In Section~\ref{sec:speedup}, we take
an orthogonal approach and analyze common and unique ingredients of
the algorithms, discussing which core ideas have been most
successful. This is followed by a discussion on closely related
problem variants and current trends in Section~\ref{sec:open},
and a conclusion in Section~\ref{sec:conclusion}.

%% file: problem.tex
\section{The Art Gallery Problem}
\label{sec:problem}

Before discussing \agp in detail, let us give a formal definition and
introduce the necessary notation.

\subsection{Problem and Definitions}\label{sec:problem.defs}

We are given a polygon $P$, possibly with holes, in the plane with vertices $V$ and $|V| = n$.
$P$ is \emph{simple} if and only if its boundary, denoted by $\partial P$, is connected.
For $p \in P$, $\vis{p} \subseteq P$ denotes all points \emph{seen} by $p$, referred to as the \emph{visibility region} of $p$, \ie, all points $p' \in P$ that can be connected to $p$ using the line segment $\overline{pp'} \subset P$.
We call $P$ \emph{star-shaped} if and only if $P = \vis{p}$ for some $p \in P$, the set of all such points $p$ represents the \emph{kernel} of $P$.
For any $G \subseteq P$, we denote by $\vis{G} = \bigcup_{g \in G} \vis{g}$.
A finite $G \subset P$ with $\vis{G} = P$ is called a \emph{guard set} of $P$; $g \in G$ is a guard.
We say that $g$ \emph{covers} all points in $\vis{g}$.
The \agp asks for such a guard set of minimum cardinality.

Note
that visibility is symmetric, i.e., $p\in \vis{q}\iff q\in
\vis{p}$. The inverse of $\vis{\cdot}$ describes all points that can
see a given point $p$. This is easily confirmed to be
\[ \vissym^{-1}(p) := \{q\in P: p\in\vis{q}\} = \vis{p} \;.\]

We use two terms to refer to points of $P$, making the
discussion easier to follow. We call a point a {\em guard
position} or \emph{guard candidate} when we want to stress its role to be selected as part of
a guard set. The second term comes from the fact that in a feasible
solution, every point in $w \in P$ needs to be covered by some visibility
polygon. We refer to such a point as {\em witness} when we use it
as certificate for coverage.

Let $G, W \subseteq P$ be sets of guard candidates and witnesses such that $W \subseteq \vis{G}$.
The \agp variant were $W$ has to be covered with a minimum number of guards, which may only be picked from $G$, can be formulated as \cidC{an} \ilp:
\begin{alignat}{3}
  \AGP(G,W)&:=\quad&\text{\makebox[3em][l]{min}}  & \sum_{g\in G} x_g \label{eq.infip.obj}\\
  &&\text{\makebox[3em][l]{s.t.}} & \sum_{g\in\vis{w}\cap G} x_g \geq 1, &\;\;& \forall w\in W,\label{eq.infip.cover}\\
  &&& x_g \in \{0,1\},            &    & \forall g\in G.\label{eq.infip.bin}
\end{alignat}
\cidC{Essentially the model above casts the  AGP variant in terms of a
  Set Covering Problem (SCP). But note}
that, depending on the choice of $G$ and $W$, $\AGP(G,W)$ may have an infinite number of variables and/or constraints, \ie, be a semi- or doubly-infinite \ilp.
We discuss three major variants of \agp:
\begin{itemize}
\item The classic \agp definition, allowing for arbitrary {\em point
    guards}, \ie, allowing to place guards anywhere within $P$. It
  requires that all of $P$, boundary and interior, is guarded. This corresponds
  to $\AGP(P,P)$. We refer to this variant as ``the'' \agp.
\item In $\AGP(V,P)$, all of $P$ has to be guarded, but guards are restricted to be placed on vertices
  of $P$ only. We refer to such guards as {\em vertex
    guards}. Trivially, a vertex guard solution is a solution for \agp
  as well, but the reverse is not necessarily true, see Figure~\ref{fig:godfried}.
\item The variant that Victor Klee actually described, \ie, where
  only the polygon's boundary needs to be guarded, is described by
  $\AGP(P,\partial P)$.
  A solution for $\AGP(P,P)$ also solves $\AGP(P,\partial  P)$, but not vice versa (see Figure~\ref{fig:wall-vs-interior}).
\end{itemize}
There are many more \agp variants that deserve (and received)
attention, however, these are the three versions that are mostly
relevant for this paper.

In the following, unless explicitly stated otherwise, we use $G$ and $W$ to indicate discretized versions of the \agp.
For example $\AGP(G,P)$ may refer to a (sub-)problem where all of $P$ needs to be guarded, but a \emph{finite} set of guard candidates is already known.
Analogously, $\AGP(P,W)$ is the version where only a finite set of points needs to be covered, and $\AGP(G,W)$ is the fully discretized version.

\begin{figure}\centering
  \def\svgwidth{.5\textwidth}
  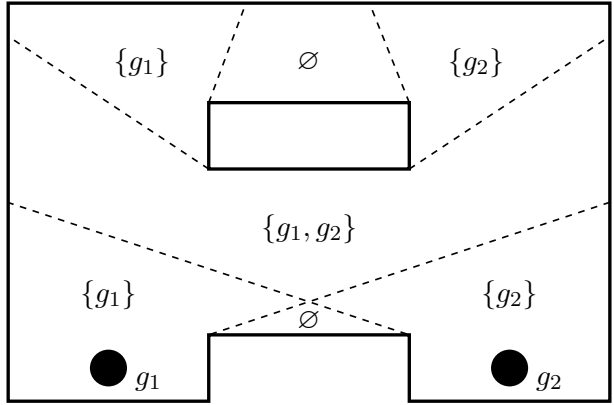
  \caption{The visibility arrangement $\visarr(\{g_1,g_2\})$ induced
    by two guards $g_1$ and $g_2$ in a polygon with one hole.}
  \label{fig:vis-arr}
\end{figure}
The semi-infinite case $\AGP(G,P)$ provides some structure that can be
exploited in algorithms. Consider Figure~\ref{fig:vis-arr}. We denote
by $\visarr(G)$ the arrangement obtained by overlaying all visibility
polygons $\vis{g}$ for every $g\in G$. Every feature (face, edge, or vertex) of
$\visarr(G)$ has a well-defined set of guards that completely sees
it.
Hence, any of those guards covers the entire feature, and we refer to
them as \avps.
We define
a partial order on them as follows: For two faces $f_1$ and $f_2$, we
define $f_1\succ f_2$ if they are adjacent in $\visarr(G)$ and the set
of guards seeing $f_1$ is a superset of those seeing $f_2$. The
maximal (minimal) elements in the resulting poset are called light
(shadow) \avps. They can be exploited to solve the two semi-infinite
cases: For given finite $G$, any subset of $G$ that covers all 
\cidC{shadow}
\avps also covers $P$, hence is feasible for $\AGP(G,P)$. For finite
$W$, there is always an optimal solution for $\AGP(P,W)$ that uses
only guards in light \avps of $\visarr(W)$, with at most one guard per \avp.

\subsection{Related Work}\label{sec:problem.rw}

Chv\'atal~\cite{c-ctpg-74} was the first to prove the famous ``Art
Gallery Theorem'', stating that $\lfloor n/3\rfloor$ guards are
sometimes necessary and always sufficient for polygons with $n$
vertices. Later Fisk~\cite{f-spcwt-78} came up with a simple proof for
this theorem, beautiful enough to be included in the
BOOK~\cite{Aigner:2009:PB:1822796}.  It also translates directly into
a straightforward algorithm to compute such a set of guards. Note that
for every $n$, there exist polygons that can be guarded by a single
point (\ie, star-shaped polygons). So any algorithm producing $\lfloor
n/3\rfloor$ guards is merely a $\Theta(n)$-approximation. There are
excellent surveys on theoretical results, especially those by
O'Rourke~\cite{o-agta-87} and Urrutia~\cite{u-agip-00} should be mentioned.

Many  variants of  the  problem have  been studied  in  the past.  For
example, Kahn \etal~\cite{kkk-tgrfw-83}  established a similar theorem
using $\lfloor  n/4\rfloor$ guards for orthogonal  polygons. There are
variants   where  the   characteristics  of   the  guards   have  been
changed. For example,  {\em edge guards} are allowed to  move along an
edge and survey all points visible to some point on this edge. Instead
of  patrolling  along  an  edge,  {\em  diagonal  guards}  move  along
diagonals,  {\em  mobile   guards}  are  allowed  to   use  both.  See
Shermer~\cite{s-rrag-92}   for   these  definitions.    Alternatively,
variations  on the  guard's task  have been  considered, for  example,
Laurentini~\cite{l-gwag-99}  required  visibility   coverage  for  the
polygon's edges only.
\cidC{Another  relevant problem  related to  the coverage  of polygons
  considers  watchman routes.   A  watchman  route is  a  path in  the
  interior of a polygon $P$ such that every point of $P$ is seen by at
  least one point in the path.  Therefore, a mobile guard moving along
  this  path  can   do  the  surveillance  of   the  entire  polygon's
  area.  Results  on  this  problem  can be  found,  for  example,  in
  Mitchell~\cite{watchman-route-soda2013}                          and
  \cite{multiple-watchman-wea2008}.}

\agp and its variants are typically hard optimization problems.
O'Rourke and Supowit~\cite{os-snpdp-83} proved \agp to be NP-hard by a
reduction from 3SAT, for guards restricted to be located on vertices
and polygons with holes. Lee and Lin~\cite{ll-ccagp-86} showed
NP-hardness also for simple polygons. This result was extended to
point guards by Aggarwal~\cite{o-agta-87}. Schuchardt and
Hecker~\cite{sh-tnhag-95} gave NP-hardness proofs for rectilinear
simple polygons, both for point and vertex guards.  Eidenbenz
\etal~\cite{esw-irgpt-01} established lower bounds on the achievable
approximation ratio. They gave a lower bound of $\Omega(\log n)$ for
polygons with holes. For vertex, edge and point guards in simple
polygons, they established APX-hardness.  For restricted versions,
approximation algorithms have been presented.
Efrat and  Har-Peled~\cite{eh-ggt-06} gave a  randomized approximation
algorithm  with  logarithmic   approximation  ratio  \cidC{for  vertex
  guards}.
Ghosh~\cite{g-aaagp-10}  presented  algorithms  for  vertex  and  edge
guards only,  with an approximation  ratio of $O(\log n)$.   For point
guards                  Nilsson~\cite{n-agmrp-05}                 gave
$O(\mathrm{OPT}^2)$-approximation algorithms  for monotone  and simple
rectilinear polygons.
\cidC{Also       for       point      guards,       Deshpande       et
  al.~\cite{deshpande-wads2007}  proposed  one  of  the  few  existing
  approximation algorithms which  is not constrained to  a few polygon
  classes. }
See Ghosh~\cite{g-aaagp-10}  for an overview of
approximation algorithms for the \agp.
\cidC{The first known exact  algorithm for point guard problem was
  proposed by Efrat and Har-Peled ~\cite{eh-ggt-06} and has complexity
  $O((nc)^{3(2c+1)})$, where $c$ is the  size of the optimal solution.
  No experimental  results with this  algorithm have been  reported so
  far.  The  exponential grow  of the running  time with  $c$ probably
  makes it useless to solve large non-trivial instances.}

%% file: vis-arr.pdf_tex
\begingroup%
  \makeatletter%
  \providecommand\color[2][]{%
    \errmessage{(Inkscape) Color is used for the text in Inkscape, but the package 'color.sty' is not loaded}%
    \renewcommand\color[2][]{}%
  }%
  \providecommand\transparent[1]{%
    \errmessage{(Inkscape) Transparency is used (non-zero) for the text in Inkscape, but the package 'transparent.sty' is not loaded}%
    \renewcommand\transparent[1]{}%
  }%
  \providecommand\rotatebox[2]{#2}%
  \ifx\svgwidth\undefined%
    \setlength{\unitlength}{144.8bp}%
    \ifx\svgscale\undefined%
      \relax%
    \else%
      \setlength{\unitlength}{\unitlength * \real{\svgscale}}%
    \fi%
  \else%
    \setlength{\unitlength}{\svgwidth}%
  \fi%
  \global\let\svgwidth\undefined%
  \global\let\svgscale\undefined%
  \makeatother%
  \begin{picture}(1,0.66850829)%
    \put(0,0){\includegraphics[width=\unitlength]{vis-arr.pdf}}%
    \put(0.50061726,0.27841119){\color[rgb]{0,0,0}\makebox(0,0)[b]{\smash{$\{g_1,g_2\}$}}}%
    \put(0.49922205,0.55606366){\color[rgb]{0,0,0}\makebox(0,0)[b]{\smash{$\varnothing$}}}%
    \put(0.83149171,0.1685791){\color[rgb]{0,0,0}\makebox(0,0)[b]{\smash{$\{g_2\}$}}}%
    \put(0.77624309,0.55531943){\color[rgb]{0,0,0}\makebox(0,0)[b]{\smash{$\{g_2\}$}}}%
    \put(0.16850829,0.1685791){\color[rgb]{0,0,0}\makebox(0,0)[b]{\smash{$\{g_1\}$}}}%
    \put(0.22375691,0.55531943){\color[rgb]{0,0,0}\makebox(0,0)[b]{\smash{$\{g_1\}$}}}%
    \put(0.5,0.12438021){\color[rgb]{0,0,0}\makebox(0,0)[b]{\smash{$\varnothing$}}}%
    \put(0.23480663,0.03045755){\color[rgb]{0,0,0}\makebox(0,0)[b]{\smash{$g_1$}}}%
    \put(0.89779006,0.03045755){\color[rgb]{0,0,0}\makebox(0,0)[b]{\smash{$g_2$}}}%
  \end{picture}%
\endgroup%

%% file: time.tex
\section{Timeline}
\label{sec:time}

After receiving mainly a theoretical treatment for over thirty years, several
groups have started working on solving the \agp using
the Algorithm Engineering methodology, aiming at providing efficient
implementations to obtain optimal, or near-optimal,
solutions.

Especially two groups, the Institute of Computing at the
University of Campinas, Brazil, and the Algorithms Group at TU
Braunschweig, Germany, developed a series of algorithms that
substantially improve in what kind of instances can be solved
efficiently. In this section, we give a chronological overview on
these efforts, and describe the algorithms that were developed.
\cidC{It should be noted that all these approaches follow similar core
  ingredients, e.g.,  the \agp  is treated as  an infinite  \scp.  As
  finite \scp instances can be solved reasonably fast in practice, the
  \agp  is  reduced  to  finite sets,  and  different  techniques  are
  employed to connect the finite and infinite cases.}\todo{R2.G1}

\input{time-stonybrook2007.tex}

\input{time-campinas2007.tex}

\input{time-torino2008.tex}

\input{time-campinas2009.tex}

\input{time-bs2010.tex}

\input{time-torino2011.tex}

\input{time-bs2012.tex}

\input{time-campinas2013.tex}

\input{time-campinas2013-full.tex}

\input{time-bs2013.tex}

\input{time-campinas2013-cur.tex}

%% file: time-stonybrook2007.tex
\subsection{Stony Brook 2007:\\ $\AGP(P,P)$ heuristics, tens of vertices}
\label{amit}

Amit \etal~\cite{amp-lgvcp-07} were among the first to experiment
with a solver for $\AGP(P,P)$, see the journal
version~\cite{amp-lgvcp-10} and the PhD thesis by
Packer~\cite{p-rgcovc-08} for extended presentations.

In this work, greedy algorithms are considered, following the same
setup: A large set $G$ of guard candidates is constructed, with the
property that $P$ can be guarded using $G$. Algorithms pick guards one
after the other from $G$, using a priority function $\mu$, until $P$ is
fully guarded. Both $G$ and $\mu$ are heuristic in nature. The authors
present 13 different strategies (i.e., choices for $G$ and $\mu$), and
identify the three that are the best: In $A_1$, $G$
consists of the polygon vertices, and of one additional point in every
face of the arrangement obtained by adding edge extensions to the
polygon. Priority is given to guards that can see the most of the
currently unguarded other positions in $G$. The second strategy,
$A_2$ follows the same idea. Additionally, after selecting a
guard $g$, it adds $\vis{g}$ to the arrangements and creates
additional candidate positions in the newly created faces. Finally,
$A_{13}$ employs a weight function $\omega$ on $G$, used as a random
distribution. In each step, a point from $G$ is selected following
$\omega$. Then, a random uncovered point $p$ is generated, and all guard
candidates seeing $p$ get their weight doubled.

To produce lower bounds, greedy heuristics for independent witnesses
(i.e., witnesses whose visibility regions do not overlap) are
considered. Using a pool of witness candidates, consisting of the
polygon's convex vertices and points on reflex-reflex edges, a witness
set is constructed iteratively. In every step, the witness seeing the
fewest other witness candidates is added, and dependent candidates are removed.

The authors conducted experiments with 40 input sets, including
randomly generated as well as hand-crafted instances, with up to 100
vertices. Both simple polygons and ones with holes are
considered. By comparing upper and lower bounds, it was found that the
three algorithms mentioned above always produced solutions that are at
most a factor 2 from the optimum. Algorithm $A_1$ was most successful
in finding optimal solutions, which happened in 12 out of 37 reported cases.

%% file: time-campinas2007.tex
\subsection{Campinas 2007:\\
 $\AGP(V,P)$ for orthogonal simple polygons, hundreds of vertices}
\label{campinas2007}

In 2007, Couto \etal~\cite{csr-eeaoagp-07,csr-eeeaoagp-08} focused on the development of an exact algorithm for the \agp with vertex guards, $\AGP(V,P)$, restricted to orthogonal polygons without holes.
To the best of our knowledge, these works were the first in the  literature  to report  extensive  experimentation  with an  exact algorithm for a variant of the \agp.
Early attempts to tackle the orthogonal $\AGP(V,P)$ also involved reductions to the \scp~\cite{erdem-paper-1,erdem-paper-2} and aimed either to obtain heuristic solutions or to solve it exactly~\cite{tomas-paper-1,tomas-paper-2}.
However, experiments in these works only considered a few instances of limited sizes.
In  contrast, in the work of Couto~\etal, thousands  of instances, some  of which  with 1000~vertices, were  tested and later  assembled into a benchmark, made publicly  available  for future comparisons~\cite{art-gallery-instances-page}, containing  new classes  of polygons including  some very
hard problem instances.

Moreover, in~\cite{csr-eeeaoagp-08}, the group in Campinas derived theoretical results that were later extended and proved to be instrumental to obtain exact solutions for more general variants of the \agp.
They showed that $\AGP(V,P)$ can be solved through a single instance of the \scp by replacing the infinite set of points $P$ by a finite set of suitably chosen witnesses from $P$.

The basic idea of the algorithm is to select a discrete set of points $W$ in $P$ and then solve the \agp variant whose objective consists in finding the minimum number of vertices sufficient to cover all points in $W$.
This discretized \agp is then reduced to an \scp instance and modeled as an \ilp.
The resulting formulation is subsequently solved using an \ilp solver, in their case, XPRESS.
If the solution to the discretized version covers the whole polygon, then an optimal solution has been found.
Otherwise, additional points are added to $W$ and the procedure is iterated.
The authors prove that the algorithm converges in a polynomial number of iterations, $O(n^3)$ in the worst case.

An important step of this exact algorithm is to decide how to construct the set of witnesses $W$. Couto \etal\ study various alternatives and investigated the impact on the performance of the algorithm.
In the first version~\cite{csr-eeaoagp-07}, a single method for selecting the initial discretization is considered, which is based on the creation of a regular grid in the interior of $P$.
In the journal version~\cite{csr-eeeaoagp-08}, four new discretizations are proposed:
\emph{Induced grid} (obtained by extending the lines of support of the edges of the polygon),
\emph{just vertices} (comprised of all vertices of $P$),
\emph{complete \avp} (consisting of exactly one point in the interior of each \avp),
and \textit{reduced \avp} (formed by one point from each shadow \avp).
The authors prove that, with the \emph{shadow \avp} discretization, that it takes the algorithm only one iteration to converge to an optimal solution for the orthogonal \agp.

The first experimental results were largely surpassed by those reported in the journal version.
Besides introducing new discretizations, the \emph{shadow \avp} discretization increased the polygon sizes fivefold (to 1000 vertices).
In total, almost 2000~orthogonal polygons were tested, including von Koch polygons, which give rise to high density visibility arrangements and, as a consequence, to larger and harder to solve \scp instances.

The authors highlight that, despite the fact that the visibility polygons and the remaining geometric operations executed by the algorithm can be computed in polynomial time, in practice, the preprocessing phase (i.e. geometric operations such as visibility polygon computation) is responsible for the majority of the running time.
At first glance, this is surprising since the \scp is known to be NP-hard and one instance of this problem has to be solved at each iteration.
However, many \scp instances are easily handled by modern \ilp solvers, as is the case for those arising from the \agp.
Furthermore, the authors also observe that, when reasonable initial discretizations of the polygon are used, the number of iterations of the algorithm is actually quite small.

Knowing that the \emph{reduced \avp} discretization requires a single iteration, albeit an expensive one timewise, the authors remark that a trade-off between the number of iterations and the hardness of the \scp instances handled by the \ilp solver should to be sought.
Extensive tests lead to the conclusion that the fastest results were achieved using the \emph{just vertices} discretization since, although many more iterations may be required, the \scp instances are quite small.

%% file: time-torino2008.tex
\subsection{Torino 2008:\\ $\AGP(P,\partial P)$, hundreds of vertices}
\label{torino2008}

In 2008, Bottino and Laurentini~\cite{bl-nospabc-08} proposed a
new algorithm for the \agp variant whose objective consists in only covering
the edges of a polygon $P$, denoted $\AGP(P,\partial P)$.
Hence, in  this version, coverage of the
interior of $P$ is not required.
Despite being  less constrained than the  original \agp, the
$\AGP(P,\partial P)$ was proven to be NP-hard~\cite{l-gwag-99}.
In  this  context,  the  authors presented an  algorithm  capable  of
optimally solving  $\AGP(P,\partial P)$  for polygons with  and without
holes provided the method converges in a finite number of steps.
This represents  a significant improvement in the  search for optimal
solutions for the \agp.

The algorithm by Bottino and Laurentini works iteratively.
First, a lower bound specific for $P$ is computed.
The  second step  consists of  solving an  instance of  the  so called
Integer Edge Covering Problem (IEC). In this problem, the objective is
also  to  cover  the  whole  boundary  of the  polygon with  one
additional restriction:  each edge must  be seen entirely by  at least
one of the selected guards.
It  is  easy  to  see  that  a  solution to the IEC is also viable for
$\AGP(P,\partial P)$ and, consequently,  its cardinality is
an upper bound for the latter.
After obtaining a viable solution,  the gap between the upper
and  lower  bounds  is checked.   If  it  is  zero  (or less  than  a
predefined threshold)  the execution  is halted.
Otherwise, a method is used to find {\em indivisible} edges, which are edges
that are entirely observed by one guard in some or all optimal solutions of
$\AGP(P,\partial P)$.  The identification of these edges can be done in
polynomial time from the visibility arrangement.
After  identifying  them, those  classified  as  not
indivisible   are    split  and   the   process   starts
over.

Tests were  performed   on  approximately  400  random polygons
with up  to 200 vertices. The instances  were divided into four classes:
simple, orthogonal, random polygons with holes and random orthogonal polygons
with holes.
Reasonable optimality percentages were obtained using  the method.  For
instance, on random  polygons with holes, optimal  results were achieved for
$65\%$ of the instances with 60 vertices.
In cases where  the program did not
reach an  optimal solution (due to the optimality gap threshold or
to timeout limits), the final  upper bound was, on average, very close
to the lower bound computed by the algorithm.
On average, for all classes of polygons, the upper bound exceeded the
lower bound by $\sim 7$\%.

%% file: time-campinas2009.tex
\subsection{Campinas 2009:\\
$\AGP(V,P)$ for simple polygons, thousands of vertices}
\label{campinas2009}

Couto \etal\ went on to study how to increase the efficiency of the
algorithm for $\AGP(V,P)$ proposed in the works discussed in
Section~\ref{campinas2007}.
A complete description of their findings can be found in a 2011
paper~\cite{crs-exmvg-11}, with a preliminary version available as a
2009 technical report~\cite{TR-IC-09-46}.
The basic steps of the algorithm are explained
in~\cite{crs-ipsagp-09}, and are illustrated in the companion video
~\cite{crs-ipsagpv-09}.

Compared to the previous works by the same authors, the new
algorithm was extended to cope with more general classes of polygons,
still without holes, but now including non-orthogonal polygons.
Experiments on thousands of instances confirmed the robustness of the
algorithm.
A massive amount of data was subsequently made publicly available
containing the entire benchmark used for these tests, see also
Section~\ref{sec:agplib}.

Essentially, some implemented procedures were improved
relative to the approach in~\cite{csr-eeeaoagp-08} to enable
handling non-orthogonal polygons.
Moreover, two new initial discretization techniques were considered.
The first one, called \emph{single vertex},
consists in the extreme case
where just one vertex of the polygon forms the initial
discretized set $W$.
As the second strategy, named \emph{convex vertices}, $W$ comprises all
convex vertices of $P$.

The authors made a thorough analysis of the trade-off
between the number and nature of the alternative discretization methods
and the number of iterations.
Their tests were run on a huge benchmark set of more than
ten thousand polygons with up to~2500 vertices.
The conclusion was that the decision over the best discretization
strategy deeply depends on the polygon class being solved.
As anticipated, the fraction of time spent in the preprocessing phase
was confirmed to be large for sizable non-orthogonal polygons and
even worse in the case of von Koch and random von Koch polygons.
Moreover, while using shadow \avps as the initial discretization
produces convergence after just one iteration of the algorithm,
the resulting discretization set can, in this case, be so large
that the time cost of the preprocessing phase overshadows the
solution of the ensued \scp.
For this reason, the \emph{just vertices} strategy
lead to the most efficient version of the algorithm, in practice,
as the small \scp instances created counterbalanced the larger
number of iterations for many polygon classes.

%% file: time-bs2010.tex
\subsection{Braunschweig 2010:\\
  Fractional solutions for $\AGP(P,P)$, hundreds of vertices}
\label{sec:bs2010}

In 2010, Baumgartner \etal~\cite{bfks-esbgagp-10} (see Kr\"oller
\etal~\cite{kbfs-esbgagp-12} for the journal version) presented an
exact algorithm for the fractional variant of $\AGP(P,P)$. In it,
solutions may contain guards $g\in P$ with a fractional value for
$x_g$. This corresponds to solving a \lp, namely the \lp relaxation of $\AGP(P,P)$ which is obtained
by replacing Constraint~\eqref{eq.infip.bin} of $\AGP(G,W)$ with
\begin{equation}
  0\leq x_g\leq 1 \quad \forall g\in G.
\end{equation}
We denote by $\AGPFrac(G,W)$ the \lp relaxation of
$\AGP(G,W)$.
Note that this is
the first exact algorithm where neither guard nor witness positions are
restricted.

The authors present a primal-dual approach to
solve the problem. They notice that the fractional $\AGPFrac(G,W)$ can be easily
solved using an \lp solver, provided $G$ and $W$ are finite and
not too large. The proposed algorithm picks small, carefully
chosen sets for $G$ and $W$. It then iteratively extends them using
cutting planes and column generation:
\begin{description}
\item [Cutting Planes]
  If there is an uncovered point
  $w\in P\wo W$, this corresponds to a violated constraint of
  $\AGPFrac(P,P)$, so $w$ is added to $W$. Otherwise the current
  solution is feasible for $\AGPFrac(G,P)$, and hence an upper bound of $\AGPFrac(P,P)$.
  We also refer to this part as {\em primal separation}.
\item [Column Generation]
  A violated constraint of the dual of $\AGPFrac(P,P)$
  corresponds to a guard candidate $g\in P\wo G$
  that improves the current solution, and $g$ is added
  to $G$.
  Otherwise the current solution optimally guards the
  witnesses in $W$, \ie is optimal for $\AGPFrac(P, W)$, and hence provides a lower bound for $\AGPFrac(P,P)$.
  We also refer to this part as {\em dual separation}.
\end{description}
It can be shown that, if the algorithm converges, it produces an
optimal solution for $\AGPFrac(P,P)$.
Furthermore, the authors use the algorithm for the integer
\agp, but only insofar that \lp solutions sometimes are integer by chance,
but without any guarantee.

The algorithm has many heuristic ingredients, e.g., in the
choice of initial $G$ and $W$ and the placement strategy for new
guards and witnesses.
The authors conducted an exhaustive experiment, with
150 problem instances with up to 500 vertices.
They compared different strategies for all
heuristic ingredients of the algorithm.
There were four separation strategies:
(1) Focusing on upper bounds by always running primal separation,
but dual only when primal failed.
(2) Focusing on lower bounds, by reversing the previous one.
(3) Always running both separators, in the hope of quickly finishing.
(4) Alternating between foci, by running primal separation
until an upper bound is found, then switching to running
dual separation until a lower bound is found, and repeating.
There were four different separators, i.e., algorithms to
select new candidates for $G$ resp.~$W$; these included
selecting the point corresponding to a maximally violated constraint,
selecting points in all \avps, a greedy strategy to find
independent rows (columns) with a large support, and placing
witnesses on \avp edges. For initial choice of $G$ and $W$,
four heuristics were used:
(1) Using all polygon vertices,
(2) starting with an empty set (for implementation reasons,
a single point had to be used here),
(3) selecting half the vertices to keep the set smaller
but still allowing full coverage, and finally two
strategies based on the work by Chwa \etal~\cite{cbkmos-gaggw-06}.
Here, $G$ is initialized to use all reflex vertices, and $W$
is initialized to have a witness on every polygon edge that is
incident to a reflex vertex.

The trial consisted of over 18,000 runs, allowing for a
direct comparison of individual parameter choices.
It was found that many instances could be solved optimally within 20~minutes.
This happened for 60\% of the 500-vertex polygons, and 85\%
of the 100-vertex polygons.
Other findings included the importance of the initial solution,
where the best strategy
(the Chwa-inspired one) led to an overall speedup factor of~2.
The best primal and dual separators were identified in a similar
 way. Furthermore, the authors were first to observe the
bathtub-shaped runtime distribution that is still prominent
in today's algorithms: Either the algorithm finishes very
quickly, usually in the first few seconds, with an optimal solutions,
or it engages in an excruciatingly slow process to find
good guards and witnesses, often failing to finish within time.

%% file: time-torino2011.tex
\subsection{Torino 2011: \\
$\AGP(P,P)$, tens of vertices}
\label{torino2011}

In 2011,  Bottino  et al.~\cite{bl-noaciagp-11}  improved
their previous work (see Section~\ref{torino2008}) by applying similar
ideas to solve the original \agp rather than the $\AGP(P,\partial P)$.
The objective  was  to  develop an  algorithm  capable  of
finding  nearly-optimal   solutions  for  full polygon coverage,
since, at  that time, there was a  lack of practical methods
for this task.

The first step  of the presented technique consists  of using the
algorithm discussed in Section~\ref{torino2008},
which allows for obtaining a lower bound and also multiple optimal solutions
for the $\AGP(P,\partial P)$.
These are then tested in the
search  for a coverage  of the  entire polygon.
According to the authors, if such solution exists, it is
automatically a nearly optimal one for $\AGP(P,P)$.
If a viable solution is not among those, guards are then
added using a greedy strategy until a feasible
solution is found.

It should be noted that there are worst-case instances for $\AGP(P,P)$
that only possess a single optimal solution, where no characterization
of the guard positions is known. Therefore this algorithm, and
none of the subsequent ones presented in this paper, can guarantee to
find optimal solutions. This common issue is discussed in more detail in Section~\ref{sec:open.degen}.

For the experiments presented in~\cite{bl-noaciagp-11}, 400
polygons
with sizes ranging from 30 to 60 vertices were examined.
As in the previous work, the following classes were tested: simple,
orthogonal, random polygons with holes and also random orthogonal
polygons with holes.
Guaranteed  optimal  solutions  were  found  in about  $68\%$  of  the
polygons tested.
Note that, in about  $96\%$ of the cases,
the solution found for $\AGP(P, \partial  P)$ in the first step of this
algorithm was also viable for $\AGP(P,P)$.
In addition, the authors also implemented the most promising techniques
by Amit \etal~(see Section~\ref{amit}), in order to enable comparison
between both works.
As a result, this technique was more successful than
the method by Amit \etal~considering the random polygons tested.

%% file: time-bs2012.tex
\subsection{Braunschweig 2012:\\
$\AGP(P,P)$, hundreds of vertices}
\label{sec:bs2012}

In 2012, the primal-dual method introduced by the Braunschweig group was extended.
The previous version could find optimal point guards, but only for the \lp
relaxation which allows fractional guards. Integer solutions could only be found by
chance. Now, two ingredients were added to find integer solutions:
An \acs{ILP}-based routine and cutting planes. See Friedrichs~\cite{f-isagplp-12}
for a detailed discussion on the cutting planes, and
Fekete~\etal~\cite{ffks-ffagp-13b,ffks-ffagp-14} for
the combined approach.
As it turned out, this algorithm could solve the classic
problem $\AGP(P,P)$ on instances of several hundreds of
vertices with holes, a factor 10 more than in previous work.

The 2012 algorithm switches between primal and dual phases.
In the primal phase, feasible solutions are sought, i.e., upper bounds.
Unlike the 2010 version, now only integer solutions are considered.
For the current set $G$ of guards and $W$ of witnesses, $\AGP(G,W)$
is solved optimally using an \ilp formulation. The visibility overlay
$\visarr(G)$ is scanned for insufficiently covered spots, and additional
witnesses are generated accordingly.
The primal phase ends when no new
witnesses are generated, with a feasible integer solution for $\AGP(G,P)$,
 and hence an upper bound for $\AGP(P,P)$. In the dual phase, new guard
positions are found using the dual arrangement $\visarr(W)$.
For that, a dual solution is needed, which is provided by solving
the \lp relaxation $\AGPFrac(G,W)$. The dual phase ends with an
optimal solution for $\AGPFrac(P,W)$, which is a lower bound for
$\AGPFrac(P,P)$, and hence also $\AGP(P,P)$. The procedure computes a narrowing sequence of upper bounds for
$\AGP(P,P)$ and lower bounds for $\AGPFrac(P,P)$, leaving the issue
of closing the integrality gap between them.
This may lead to terminating with a suboptimal solution, however with
a provided lower bound.
As a leverage against this shortcoming,
cutting planes are employed to raise the lower bounds~\cite{f-isagplp-12}.
Two classes of facet-defining inequalities for the convex hull of
all feasible integer solution of $\AGP(G,W)$ are identified.
While the NP-hardness of \agp indicates that it is hopeless to find
a complete polynomial-size facet description,
it is shown that the new inequalities contain a large set of facets,
including all with coefficients in $\{0,1,2\}$,
see also~\cite{springerlink:10.1007/BF01589093}.
The dual phase is enhanced with separation routines for the
two classes, consequently improving the lower bounds, and often
allowing the algorithm to terminate with provably optimal solutions.

To evaluate this work, experiments were conducted on four different
polygon classes, sized between 60 and 1000 vertices.
These included both orthogonal and non-orthogonal instances,
both with and without holes, and polygons where optimal solutions
cannot use vertex guards.
Different parametrizations of the algorithms were tested,
and it was found that the \acs{ILP}-based algorithm itself
(without applying cutting planes) could identify good
integer solutions, sometimes even optimal ones, and considerably
surpassed the previous 2010 version.
The algorithm was able
to find optimal solutions for 500-vertex instances quite often.
Instances with 1000 vertices were out of reach though.

%% file: time-campinas2013.tex
\subsection{Campinas 2013:\\
$\AGP(P,P)$, hundreds of vertices}
\label{campinas2013}

The work by Tozoni \etal~\cite{trs-qosagp-13}
generalizes to $\AGP(P,P)$ the ideas developed for $\AGP(V,P)$
by Couto \etal~(see Section~\ref{campinas2009}).
The paper proposes an algorithm that iteratively generates upper and
lower bounds while seeking to reach an exact solution.
Extensive experiments were carried out which comprised 1440
simple polygons with up to 1000 vertices from several
classes, all of which were solved to optimality in a matter
of minutes on a standard desktop computer.
Up to that point in time, this was the most robust and effective
algorithm available
for $\AGP(P,P)$, for simple polygons.
The restriction to simple polygons in this version as well as
earlier versions of the Campinas branch originates from the fact
that no visibility algorithm for general polygons was available to
the group in Campinas, yet.

The algorithm generates, through a number of iterations,
lower and upper bounds
for the $\AGP(P,P)$ through the resolution of the two semi-infinite discretized variants
of the original \agp, namely $\AGP(P,W)$
(asking for the minimum number of guards that are sufficient to cover the finite
set $W$ of witnesses)
and $\AGP(G,P)$
(computing the minimum number of guards from $G$ that are sufficient to cover $P$).
Notice that in these variants, either the witness or the guard
candidate set is infinite, preventing the formulation of these problem
variants as an \ilp.
However, remarkable results~\cite{trs-qosagp-13} show that
both variants can be reduced to a compact set covering problem.

To solve $\AGP(P,W)$ instance, the algorithm constructs $\visarr(W)$,
and chooses the vertices of the light \avps to become part of the guard
candidates set $G$.
Assuming that $|W|$ is bounded by a polynomial in $n$, the same
holds for $|G|$. Therefore, the \scp instance corresponding to
$\AGP(G,W)$ admits a compact \ilp model.
Tozoni \etal\ showed that an optimal solution for $\AGP(P,W)$ can be
obtained by solving $\AGP(G,W)$.
Thus, the algorithm computes a lower bound for $\AGP(P,P)$ using an \ilp
solver.

Now, to produce an upper bound for $\AGP(P,P)$, an idea similar to the
one developed by Couto \etal~\cite{crs-exmvg-11} to solve the
$\AGP(V,P)$ is used.
The procedure starts with the same sets $G$ and $W$ used for the lower
bound computation.
The $\AGP(G,W)$ is solved as before.
If the optimal solution found in this way covers $P$, then it is also
feasible for $\AGP(P,P)$ and provides an upper bound for the
problem.
Otherwise, new witnesses are added to the set $W$ and the procedure
iterates.
The upper bound procedure is known to converge in a number of
iterations that is polynomial in $n$.

The lower and upper bound procedures are repeated until the gap
between the two bounds reaches zero or a predefined time limit is
reached.
For certain initial discretization sets and strategies for
updating the witness set, one can construct fairly simple
instances that lead the algorithm to run indefinitely.
Therefore, it remains an important open question whether there
exists a discretization scheme that guarantees that the algorithm
always converges, see also Section~\ref{sec:open.degen}.

An important step of this algorithm, which greatly affects
the performance of the final program, is how the initial witness
set should be chosen and updated throughout the iterations.
Two initial discretizations were tested in practice and are worth
noting. The first one, called \textit{Chwa-Points}, is based
on the work by Chwa et al.~\cite{cbkmos-gaggw-06} and chooses the middle
points of reflex-reflex edges and the convex vertices that are
adjacent to reflex vertices.
This is similar to the initialization
used in~\cite{bfks-esbgagp-10,kbfs-esbgagp-12}.
The second, called \textit{Convex-Vertices}, comprises
all convex vertices of $P$.

The computational results obtained by this algorithm
confirmed its robustness.
A total of 1440 instances were tested from different polygon classes,
including simple, orthogonal and von Koch ones.
Optimal solutions were found for all of them.
Also, comparisons with previous published papers showed that the
algorithm was effective and far more robust than its competitors.
Experiments with different initial witness sets revealed that, on
average, \emph{Chwa Points} attained the best results. However, on von Koch
Polygons, \emph{Convex Vertices} performed better.

%% file: time-campinas2013-full.tex
\subsection{Campinas 2013 (Journal version):\\
$\AGP(P,P)$, thousands of vertices}
\label{campinas2013J}

After presenting an algorithm for \agp with point guards in
spring 2013 (see  Section~\ref{campinas2013}),  the
research group from Campinas continued working on the subject.
In  this  context, improvements  were  implemented,
including the development of their own visibility algorithm
that was also able to handle polygons with holes, giving
rise to a new version of the algorithm~\cite{DaviPedroCid-J-002013}.
The resulting implementation is  able to solve polygons with thousands
of  vertices in  a  few minutes  on  a  standard  computer.

Several major improvements were introduced in order
to reach this redesigned version.
Among them, a Lagrangian Heuristic method (see Section~\ref{sec:lagrange}) was implemented to help the
\ilp solver expedite the  computation of optimal solutions for \scp
instances.
Moreover, a procedure for removing redundant variables and constraints
from the \scp formulation was  also used to  speed up the
\ilp resolution process.

One of the most effective  changes consisted in reversing the point of
view of visibility testing from  the perspective of the guards to that
of the witnesses. Since these are fewer in number and their arrangement
has already been computed, much of the geometric computation is simplified.

In  the  end,  2440  instances  were  tested and
optimal solutions were found for more than 98\% of them.
The test bench included several different classes of polygons, with and
without holes, with up to 2500 vertices.
Besides the  classes tested in the previous version~\cite{trs-qosagp-13},
the authors also used a newly created benchmark instance for polygons with
holes (see also Section~\ref{sec:agplib}) and  the  spike  polygons
presented  by  Kr\"oller \etal~\cite{kbfs-esbgagp-12}.
Also, comparisons were made
with  the  work of Kr\"oller  et al.~and an analysis  of the
effects obtained from different discretizations for the initial
witness set were presented.
Moreover, the authors evaluated the  impact of  using a
Lagrangian   heuristic   on   the   overall   performance   of   the
method and concluded that it reduces the average
execution time in most of the cases.

%% file: time-bs2013.tex
\subsection{Braunschweig 2013 (Current version):\\
  $\AGP(P,P)$, thousands of vertices}
\label{sec:bs2013}

A deeper runtime analysis of the former code from 2012 revealed
that the main bottlenecks where the geometric subroutines, namely
(i) the computation of visibility polygons
(an implementation of a $\bigO(n \log n)$ rotational sweep
as in Asano~\cite{asano1985efficient}),
(ii) the overlays of these visibility polygons
to form $\visarr(G)$ and $\visarr(W)$ ($\bigO(n^2m^2\log(nm))$,
where $m$ is the size of $G$ resp.~$W$),
and
(iii) point location algorithms to determine membership in \avps.
This was somewhat surprising as all of these algorithms
have fairly low complexity, especially when compared to \lp
solving (worst-case exponential time
when using the Simplex algorithm) and \ilp solving
(NP-hard in general, and used to solved the NP-hard
Set Cover problem).
Still the geometric routines made up for over
90\% of the runtime.

The group in Braunschweig focused on the
improvement of these geometric subroutines:
(i)
A new \cgalcite package for visibility polygon computation
was developed in Braunschweig~\cite{cgal:visibility},
which contains a new algorithm named triangular
expansion~\cite{bhhhk-ecvo-2014}.
Though the algorithm only guarantees an $O(n^2)$ time
complexity, it usually performs several magnitudes faster
than the rotational sweep.
(ii)
The code now uses the lazy-exact kernel~\cite{pf-glese-06}, which delays
(or even avoids) the construction of exact coordinates of
intersection points as much as possible.
The impact is most evident in the
construction of the overlays, which contain many intersection
points.
(iii) The algorithm was restructured to allow a batched
point location~\cite[Sec.~3]{cgal:wfzh-a2-12}%
\footnote{This is an an $O((n+m)\log n)$ sweep line algorithm,
where n is the number of polygon vertices and m the number
of query points.}
of all already existing guards (or witnesses) with respect to a new
visibility polygon at once.

The new code now runs substantially faster, allowing it to
solve much larger instances than the previous one.
This paper contains the first experimental evaluation of this new algorithm.
Section~\ref{sec:experimental_evaluation} contains results from
running the algorithm and comparing it to the other approaches presented here.
Section~\ref{sec:speedup} discusses the speedup obtained by the new subroutines.

%% file: time-campinas2013-cur.tex
\subsection{Campinas and Braunschweig 2013 (Current version):\\
$\AGP(P,P)$, thousands of vertices}
\label{campinas2013-cur}

This implementation is the result of a joint effort by the Braunschweig and the
Campinas groups.
With the intent of achieving robustness, its core is
the algorithm from Campinas (Section~\ref{campinas2013J}),
refitted with optimizations from Braunschweig that greatly
improved its efficiency.

The new code now also uses the lazy exact kernel (cf.~\ref{sec:bs2013}) of \cgal and
the triangular expansion algorithm~\cite{bhhhk-ecvo-2014} of the new
visibility package~\cite{cgal:visibility} of \cgal.
While the impact of the new visibility polygon algorithm was
huge for both approaches the usage of the lazy kernel
was also significant since the overlays in the approach of Campinas
contain significantly more intersection points.
To see more about how changes in kernel and visibility affect the solver,
consult Section~\ref{sec:speedup}.

Moreover, the current version of Campinas also includes new
approaches on the algorithm side.
One of the ideas developed was to postpone the computation of an upper bound
(solving $\AGP(G,P)$) to the time that a good lower bound, and, consequently, a
``good'' set of guard candidates is obtained.
This can be done by repeatedly solving only $\AGP(P,W)$ instances until an
iteration where the lower bound is not improved is reached. This situation
possibly means that the value obtained will not change much in the next
iterations. It also increases the chances that the first viable solution found
is also provably optimal, which automatically reduces the number of $\AGP(G,P)$
instances which must be re-solved.

Other changes that are the inclusion of a new strategy
for guard positioning, where only one interior point from each light \avp
is chosen to be part of the guard candidate set (instead
of all its vertices), and the
possibility of using \cplexcite solver instead of XPRESS.

This new version was tested in experiments conducted for this
paper, using 900 problem instances ranging from 200 to 5000 vertices.
Section~\ref{sec:experimental_evaluation} presents the
obtained results in detail. The implementation proved to be efficient and robust for all classes of
polygons experimented.

%% file: experiments.tex
\section{Experimental Evaluation}
\label{sec:experimental_evaluation}

To assess how well the \agp can be solved using current
algorithms, and how their efficiency has developed over the last years, we
have run exhaustive experiments.
The experiments involve all algorithms for which we could get working
implementations, and were conducted on the same set of instances and on the same
machines, see Section~\ref{sec:exp-setup}.
We refrain from providing comparisons based on numbers from the literature.

We had several snapshots from the Braunschweig and Campinas code
available, these are:
\begin{itemize}
\item For Braunschweig, the versions from 2010
  (Section~\ref{sec:bs2010}), 2012 (Section \ref{sec:bs2012}), and 2013
  (\ref{sec:bs2013}).
  These will be referred to as \bsx, \bsxii, and \bscur, respectively.
\item
  For Campinas, the version from 2009 (Section~\ref{campinas2009}),
  and the two snapshots from 2013
  (Sections~\ref{campinas2013} and~\ref{campinas2013J}).
  These will be referred to as \cix, \cxiiisea and \cxiiifull, respectively.
\item
  The latest version is the combined approach from Campinas and Braunschweig that
  was
  obtained during a visit of Davi~C.~Tozoni to Braunschweig
  (Section \ref{campinas2013-cur}), which we refer to as \ccur.
\end{itemize}
The older versions have already been published, for these we
provide a unified evaluation. The versions \bscur and
\ccur\ are, as of yet, unpublished.

\subsection{AGPLib}
\label{sec:agplib}

\input{cps-agplib-description}

\begin{figure}
  \centering
  \begin{subfigure}[b]{0.3\textwidth}
    \includegraphics[width=\textwidth]{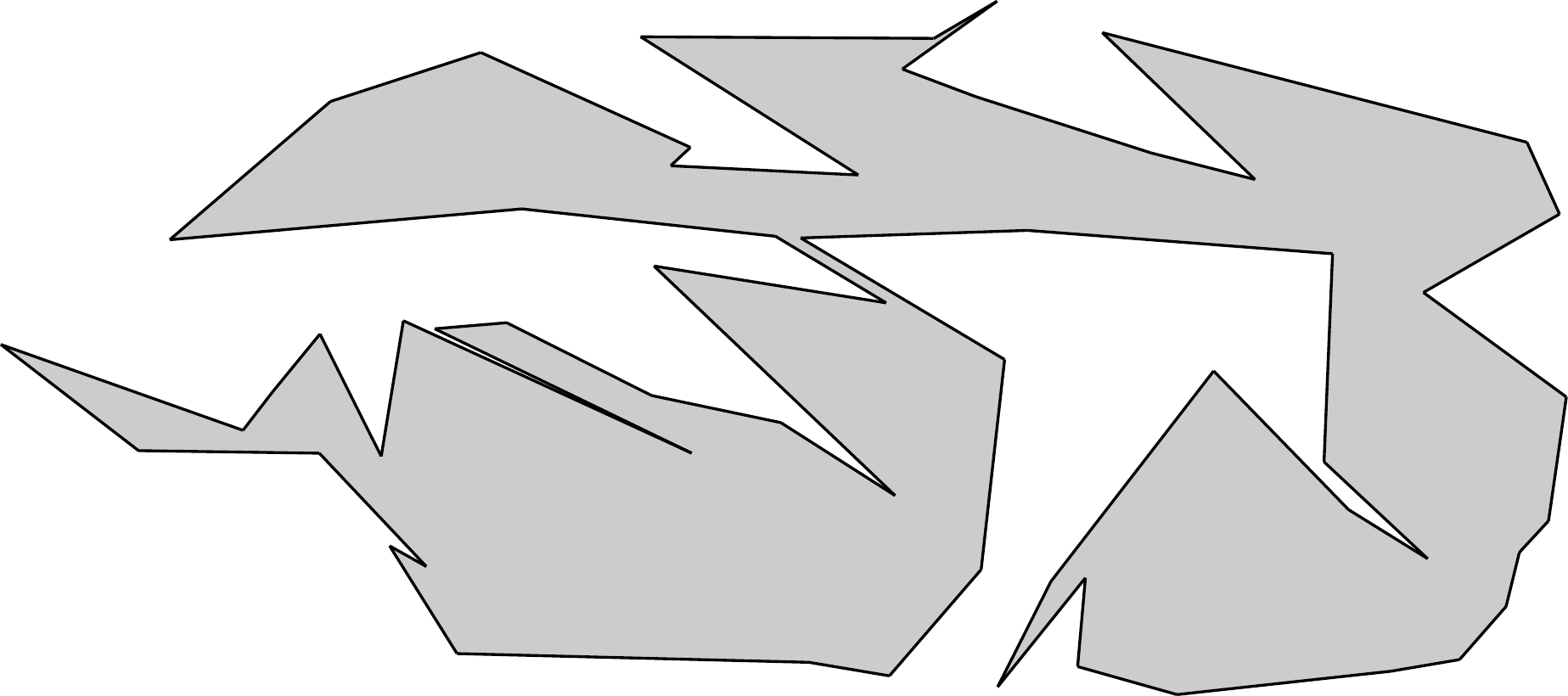}
    \caption{simple}
    \label{fig:testpolygon-simple}
  \end{subfigure}
  ~
  \begin{subfigure}[b]{0.3\textwidth}
    \includegraphics[width=\textwidth]{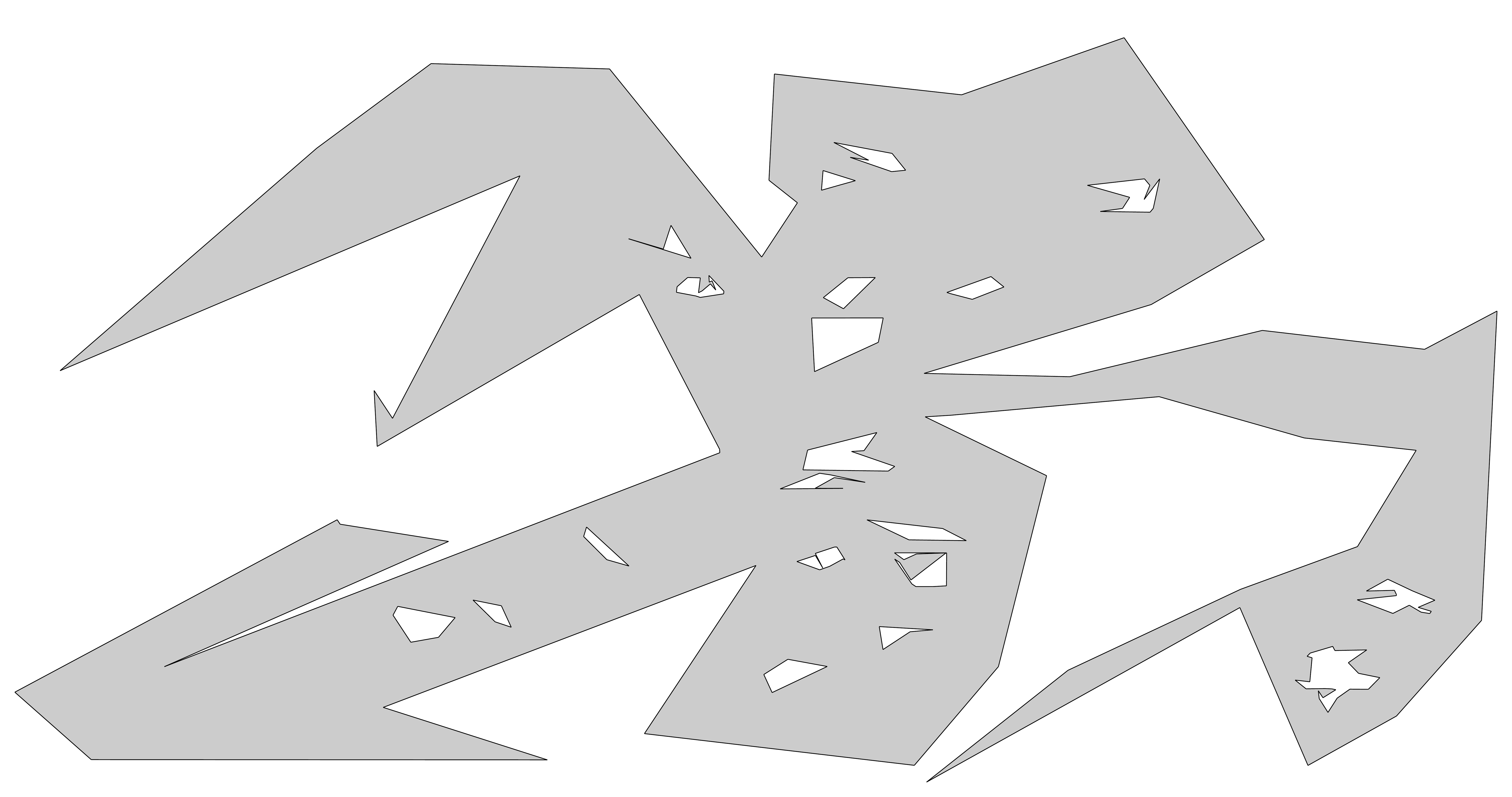}
    \caption{simple-simple}
    \label{fig:testpolygon-simple-simple}
  \end{subfigure}
  ~
  \begin{subfigure}[b]{0.3\textwidth}
    \includegraphics[width=.7\textwidth]{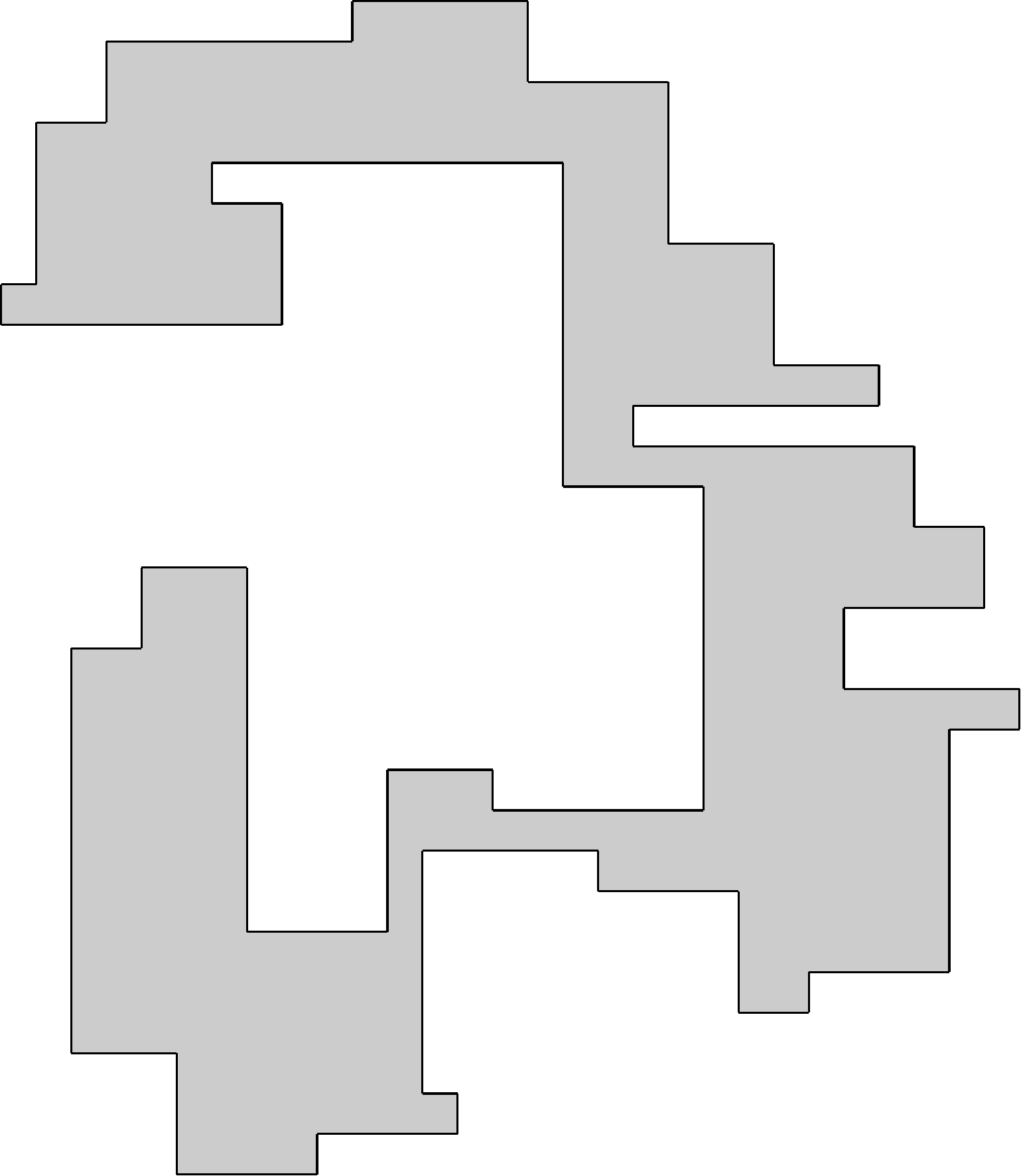}
    \caption{ortho}
    \label{fig:testpolygon-ortho}
  \end{subfigure}
  ~
  \begin{subfigure}[b]{0.3\textwidth}
    \includegraphics[width=\textwidth]{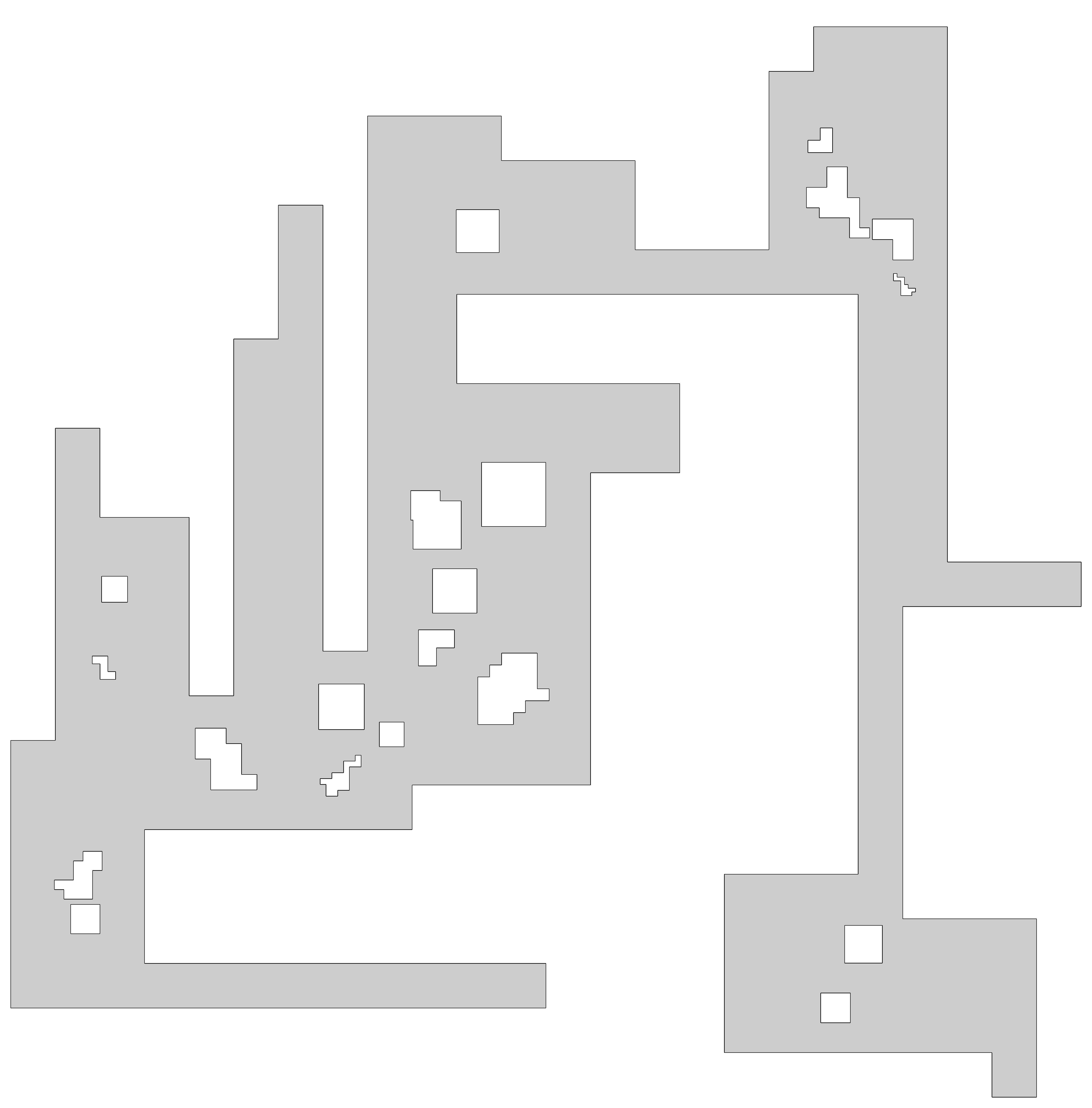}
    \caption{ortho-ortho}
    \label{fig:testpolygon-ortho-ortho}
  \end{subfigure}
  ~
  \begin{subfigure}[b]{0.3\textwidth}
    \includegraphics[width=\textwidth]{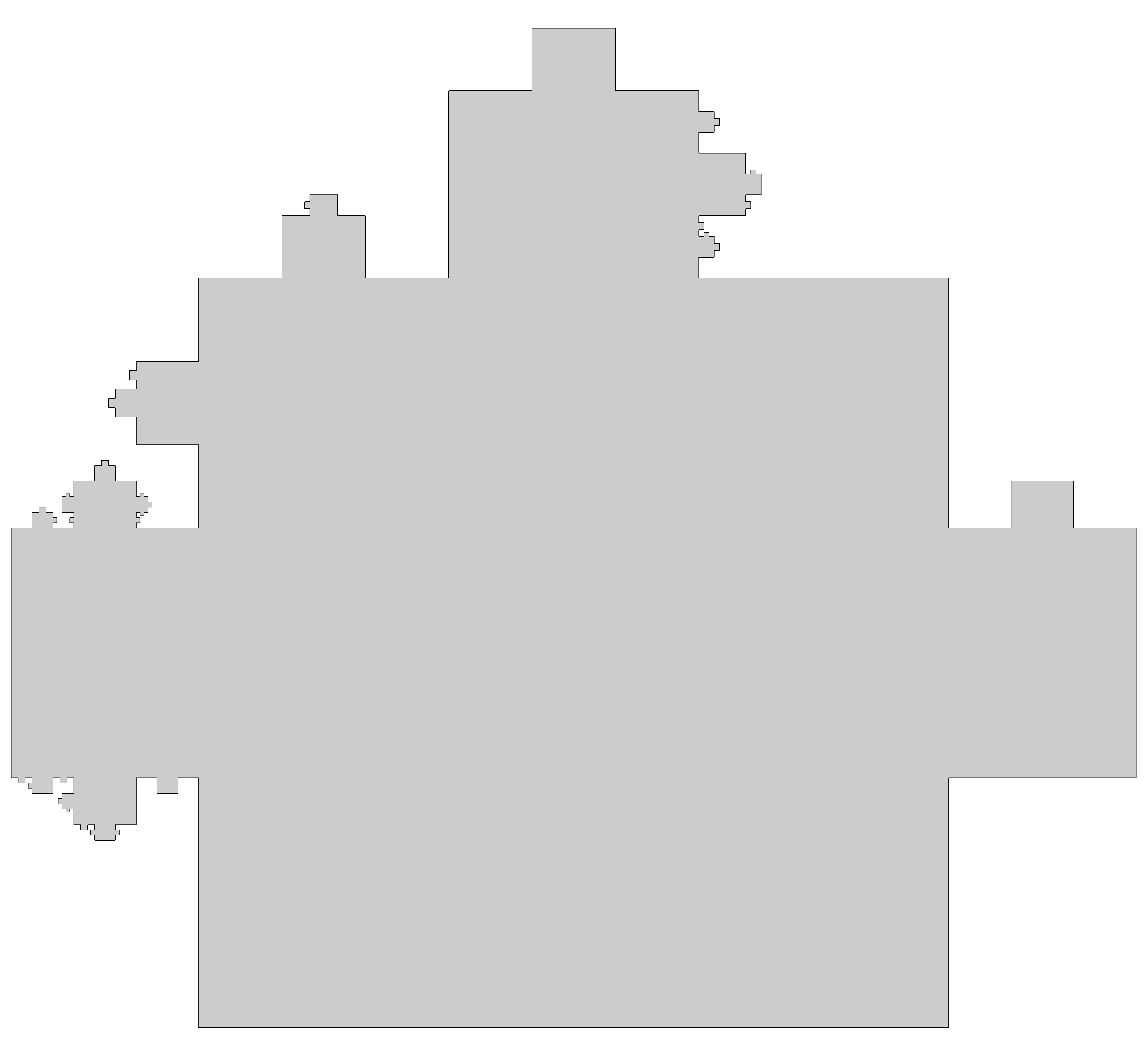}
    \caption{vonKoch}
    \label{fig:testpolygon-koch}
  \end{subfigure}
  ~
  \begin{subfigure}[b]{0.3\textwidth}
    \includegraphics[width=\textwidth]{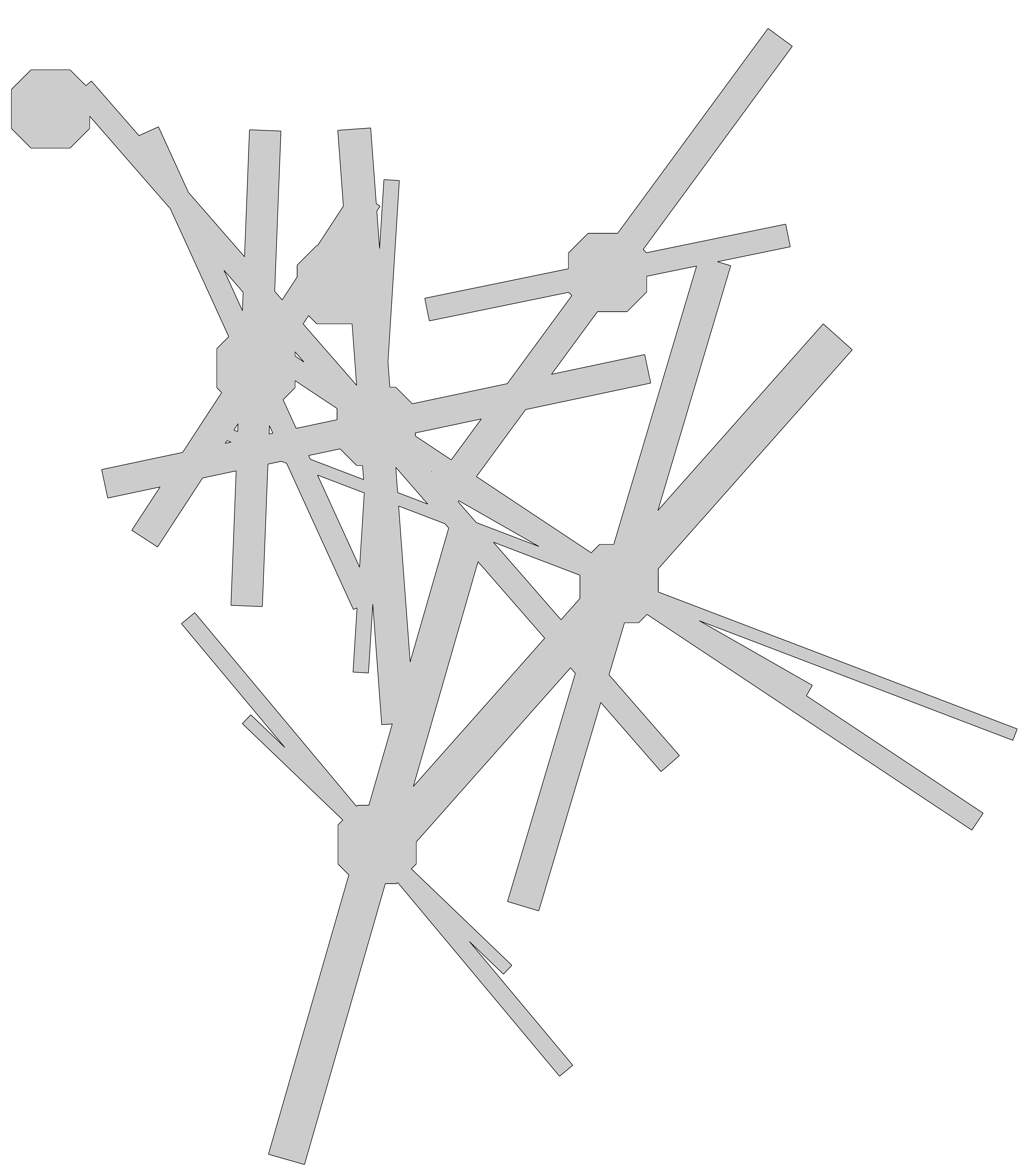}
    \caption{spike}
    \label{fig:testpolygon-spike}
  \end{subfigure}
  \caption{Example instances of different polygon classes.}
    \label{fig:testpolygon}
\end{figure}

\begin{description}
\item[``simple'']
  Random non-orthogonal simple polygons as in Figure~\ref{fig:testpolygon-simple}.
\item[``simple-simple''] Random non-orthogonal polygons as in
  Figure~\ref{fig:testpolygon-simple-simple}. These are generated like
  the ``simple'' polygon class, but with holes.  The holes are also
  generated like the first class and randomly scaled and placed until
  they are in the interior of the initial polygon.
\item[``ortho''] Random floorplan-like simple polygons with
    orthogonal edges as in Figure~\ref{fig:testpolygon-ortho}.
  \item[``ortho-ortho''] Random floorplan-like orthogonal polygons
    as in Figure~\ref{fig:testpolygon-ortho-ortho}.
    As the simple-simple class, these polygons are generated by using one polygon
    of the ortho class as main polygon, and then randomly scaling and translating
    smaller ortho polygons until they are contained within the main polygon's
    interior, where they are used as holes.
\item[``von Koch'']
	Random polygons inspired by randomly pruned Koch curves,
	see Figure~\ref{fig:testpolygon-koch}.
\item[``spike'']
	Random polygons with holes 
	as in Figure~\ref{fig:testpolygon-spike}.
	Note that this class is specifically designed to provide polygons that encourage
	placing point guards in the intersection centers.
	It has been published along with the \bsx algorithm (Section~\ref{sec:bs2010}), which was the first capable of placing point guards.
\end{description}

\ignore{
\begin{figure}[ht]
	\centering
	\subfigure{
		\includegraphics[height=2.5cm]{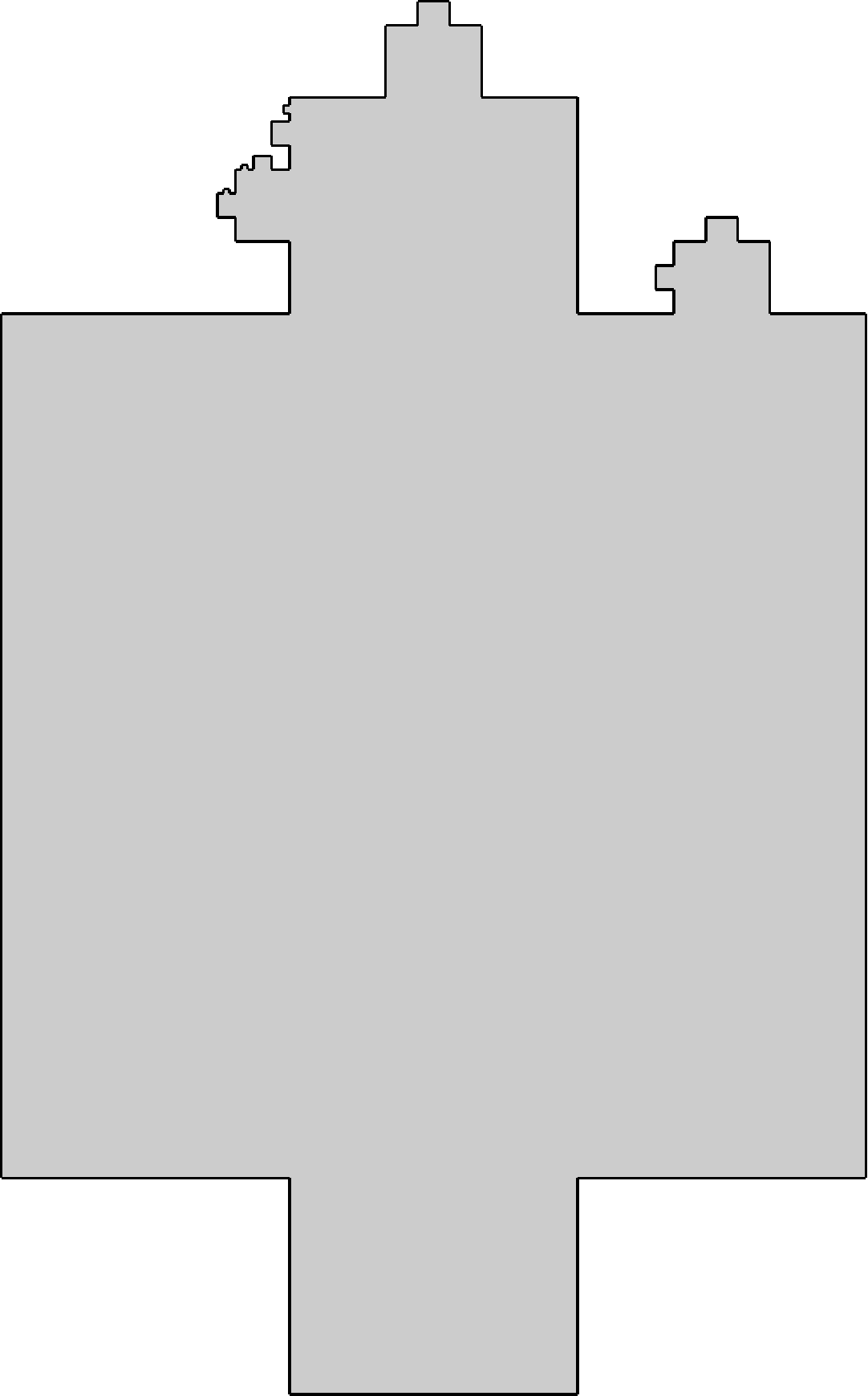}
		\label{fig:testpolygon-koch}
	}
	\subfigure{
                \includegraphics[height=2.5cm]{testpolygon_orthogonal_60.pdf}
		\label{fig:testpolygon-orthogonal}
	}
	\subfigure{
                \includegraphics[height=2.5cm]{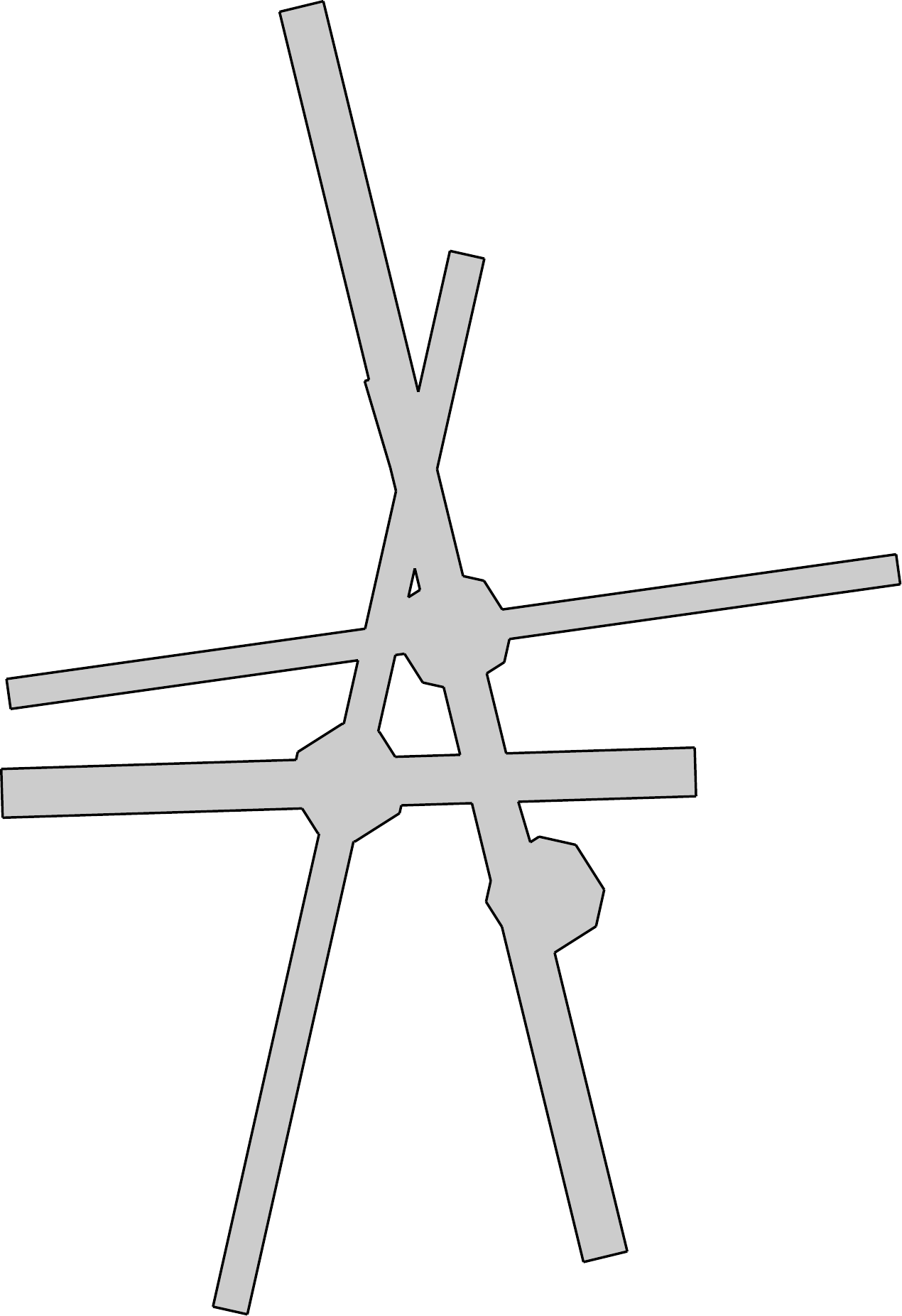}
		\label{fig:testpolygon-spike}
	}
	\subfigure{
                \includegraphics[height=2.5cm]{testpolygon_simple_60.pdf}
		\label{fig:testpolygon-simple}
	}
	\caption[Test Polygon Classes]{%
	Small \emph{von Koch}, \emph{Orthogonal}, \emph{Spike} and \emph{Simple} test polygons.
	}
	\label{fig:testpolygon}
\end{figure}
}

\subsection{Experimental Setup}
\label{sec:exp-setup}

The experiments were run on identical PCs with eight-core Intel Core i7-3770 CPUs
at 3.4\,GHz, 8\,MB cache, and 16\,GB main memory running a 64-bit Linux 3.8.0 kernel.
All algorithms used version 4.0 of \cgalcite and \cplexcite 12.5.
The only component using concurrency is the \ilp solver \cplex, everything else was single-threaded.
For each polygon-class/complexity combination, we tested 30~different polygons.
Each test run had a runtime limit of 20~minutes.

\subsection{Results}
\newlength{\sizecolwdt}\settowidth{\sizecolwdt}{2000}
\def\BX#1{\makebox[\sizecolwdt][r]{#1}}

Historically, the two lines of algorithms have been working towards
\agp from different angles.
Campinas focused on binary solutions, which initially came at the expense of being limited to given guard discretization, like vertex guards: $\AGP(V,P)$.
The Braunschweig work started with point guards, but the price were fractional solutions: $\AGPFrac(P,P)$.
It only became possible to reliably solve the binary \agp with point guards, $\AGP(P,P)$, with the \bsxii and \cxiiisea algorithms.

Therefore, we first sketch progress for the \agp with vertex guards and the fractional \agp before discussing the experimental results for \agp with point guards itself.

\subsubsection{Vertex Guards}

$\AGP(V,P)$ is one of the semi-infinite variants.
We believe it to be a considerably simpler problem than $\AGP(P,P)$ for two reasons:
(1)~Both variants are NP-hard, but we know $\AGP(V,P)$ is in NP, which is uncertain
for $\AGP(P,P)$ as it is unknown if there is a polynomial-size representation of
guard locations.
(2)~Experience and experimental results indicate that finding good guard
candidates is the hardest part of the problem and leads to many iterations;
but for $\AGP(V,P)$ we only have to place witnesses and solve \scp instances,
which is usually possible in a comparably short time frame with a good \ilp
solver.
The first experimental work on $\AGP(V,P)$ was Campinas~2007,
but unfortunately, the implementation is no longer available.

\begin{table}\centering
  \begin{tabular}{|l|c|c|c|c|c|c|c|c|c|c|}\hline
    &\multicolumn{5}{c|}{\cidC{\bf polygons without holes}}        & \multicolumn{5}{c|}{\bf polygons with holes}  \\
    &\BX{200}&\BX{500}&\BX{1000}&\BX{2000}&\BX{5000}&\BX{200}&\BX{500}&\BX{1000}&\BX{2000}&\BX{5000}\\\hline
    \cix
    &100.0&100.0&100.0&100.0&63.3&
    --&--&--&--&--
    \\
    \cxiiisea&
    100.0&100.0&100.0&100.0&63.3&
    --&--&--&--&--
    \\
    \bscur&
    100.0&100.0&100.0&100.0&88.9&
    100.0&97.8&67.8&77.8&66.7
    \\
    \cxiiifull&
    100.0&100.0&100.0&100.0&37.8&
    100.0&100.0&66.7&65.6&0.0
    \\
    \ccur&
    100.0&100.0&100.0&100.0&100.0&
    100.0&100.0&77.8&88.9&66.7
    \\\hline
    \bsx*&
    100.0&100.0&100.0&100.0&33.3&
    100.0&100.0&100.0&98.9&0.0
    \\\hline
  \end{tabular}
  \caption{
    Optimality rates for vertex guards. Notice that \bsx\
    finds fractional vertex guard solutions, whereas the others
    find integer ones.}
  \label{tab:fopt-vguard}
\end{table}

Table~\ref{tab:fopt-vguard} shows optimality rates, i.e., how many of
the instances each implementation could solve, given a 20~minute time
limit per instance.
\cidC{The  polygons  were grouped  in  two  categories: those  without
  holes, including the instances classes {\bf simple}, {\bf ortho} and
  {\bf von  Koch}, and those with  holes composed by  the instances in
  the  classes   {\bf  simple-simple},  {\bf   ortho-ortho}  and  {\bf
    spikes}.}\todo{R2.D4}
The Campinas versions prior to \cxiiifull could not deal with
holes in input polygons, so these entries are empty. It should also
be noted that \bsx\ solves the easier case of fractional vertex
guards.
It is clearly visible how all algorithms (including the five-year-old \cix)
can solve all simple polygons with up to 2000 vertices as well as
most simple 5000-vertex polygons.
For instances with holes, however, the solution percentages of all
algorithms (except \bsx which solves an easier problem) start
significantly dropping at 1000 vertices.
This demonstrates two effects: First, for smaller sizes, the problem is
easier to solve as the search for good guard candidates is
unnecessary. Second, for larger sizes, finding optimal solutions to
large instances of the NP-hard \scp dominate, resulting in a
computational barrier.
The difficulty  to handle large SCP  instances also shows up  when we
consider the results of the  Campinas codes \cxiiisea and \cxiiifull.  As
the size of the  polygons increases and the SCPs to  be solved grow in
complexity, the Lagrangian heuristic employed by \cxiiifull version uses
more  computational time  but does  not help  the ILP  solver to  find
optimal solutions for the AGP(G,W) instances, due to the deterioration
of  the primal  bounds.  This  inefficiency causes  a decrease  in the
solver's performance, as  can be seen in the optimality  rate shown in
Table  1 for  simple polygons  with 5000  vertices. In  this case,  if
\cxiiifull did not use the Lagrangian  heuristic by default, a result at
least similar to that obtained by \cxiiisea would be expected.

The high solution rates allow us to directly analyze the speedup achieved over time.
Table~\ref{tab:speedup-vguard} shows how much faster than \cix\ later algorithms could solve the problem.
The shown numbers are log-averages over the speedup against \cix for all instances solved by both versions.
As \cix\ cannot process holes, this analysis is restricted to simple polygons. It is clearly visible that
\bscur\ is about five times faster then \cix, and the changes from
\cxiiifull\ to \ccur\ led to a speedup factor of about seven. These stem
from a number of changes between versions, however, roughly a factor 5
can be attributed to improvements in geometric subroutines --- faster visibility
algorithms, lazy-exact \cgal kernel, reduced point constructions.
We discuss the influence of geometry routines in Section~\ref{sec:speedup.geometry}.

\input{tab-speedup-vguard}

\subsubsection{Fractional Guards}

The Braunschweig line of work started with solving $\AGPFrac(P,P)$, the fractional point guard variant, and all Braunschweig versions, even those designed for binary solutions, still support the fractional \agp.
Table~\ref{tab:gopt-frac} shows how often the three implementations could find optimal solutions, and how often they achieved a 5\%~gap by the end of the 20-minute runtime limit.
\cidC{Here again,  the polygons  have been grouped: those with holes
and those without holes.}\todo{R2.D4}

Unsurprisingly, there is no significant difference between the \bsx\ and \bsxii versions, the development between these snapshots focused on the integer case.
The improvements from \bsxii\ to \bscur\ stem from improved geometry subroutines which are beneficial to both, the binary and the fractional mode.
It can be seen that near-optimal solutions are obtained almost every
time, but the gap is not always closed. Furthermore, with the
20-minute time limit, there is an barrier between 2000 and 5000
vertices, where the success rate drops sharply, indicating that the
current frontier for input complexity lies roughly in this range.

\input{tab-gopt-frac}

\subsubsection{Point Guards}

We turn our attention to the classic \agp, $\AGP(P,P)$:
Finding integer solutions with point guards. We report optimality in three different ways:
Which percentage of instances could be solved optimally with a matching lower bound (i.e., proven optimality) is reported in Table~\ref{tab:opt-agp};
we show in how many percent of the cases an instance could be solved optimally,
whether or not a matching bound was found in Table~\ref{tab:opt-with-no-proof-agp};
Table~\ref{tab:opt-with-gap} reports how many percent of the solutions
were no more than 5\% away from the optimum. This allows to
distinguish between cases where \bscur\ does not converge, and cases
where the integrality gap prevents it from detecting optimality.

\input{tab-opt-agp}

\input{tab-opt-no-proof-agp}

\input{tab-gap5-agp}

The \ccur\ implementation solves the vast majority of instances from our test set to proven optimality, the only notable exception being some classes of very large polygons with holes and the 5000-vertex Koch polygons.
Given how the best known implementation by~2011, the Torino one from Section~\ref{torino2011}, had an optimality rate of about 70\% for 60-vertex instances, it is clearly visible how the developments in the last years pushed the frontier.
With \ccur, instances with 2000~vertices are usually solved to optimality, showing an increase in about two orders of magnitude.
The success of \ccur\ is multifactorial:
It contains improved combinatorial algorithms as well as faster geometry routines, most notably a fast visibility implementation.
Section~\ref{sec:speedup} discusses its key success factors.

It can be seen from Table~\ref{tab:opt-with-gap} that many algorithms are able to find near-optimal solutions (5\% gap) for most instances, indicating that for practical purposes, all 2013 algorithms perform very well.
The frontier on how large instances can be solved with small gap is between~2000 and~5000 vertices for most polygons with holes and beyond~5000 vertices for simple polygons.

Comparing Tables~\ref{tab:opt-agp}--\ref{tab:opt-with-gap}, it can be seen that the primal-dual approach (\bsx and \bsxii) produces decent upper bounds, often optimal ones, but does have an issue with finding matching lower bounds.
This drawback has been much improved in \bsxii but is still measurable.

Finally, we analyze how difficult the individual instance classes are.
In Tables~\ref{tab:opt-agp}--\ref{tab:opt-with-gap}, we group them by size and based on whether they feature holes.
Table~\ref{tab:foptinst-agp} shows optimality rates for each class.
We restrict presentation to \bscur\ here, for the simple reason that it has the highest variation in reported rates.
\begin{table}
  \centering
  \begin{tabular}{|l|rrrrr|r|}\hline
      &\BX{200}&\BX{500}&\BX{1000}&\BX{2000}&\BX{5000}&Avg.\\\hline
      Simple & 96.7 & 96.7 & 90.0 & 60.0 & 26.7  & 74.0 \\
      Orthogonal & 96.7 & 93.3 & 86.7 & 70.0 & 40.0 & 77.3 \\
      simple-simple & 86.7 & 60.0 & 13.3 & 0.0 & 0.0 & 32.0 \\
      ortho-ortho & 86.7 & 53.3 & 16.7 & 0.0 & 0.0 & 31.3 \\
      von Koch & 100.0  & 93.3 & 96.7 & 86.7 & 0.0 & 75.3 \\
      Spike & 96.7 & 100.0 & 100.0 & 100.0 & 96.7 & 98.7 \\\hline
  \end{tabular}
  \caption{Optimality rates for \bscur\ on different instance classes.}
  \label{tab:foptinst-agp}
\end{table}

In each class, we see a continuous decline with increasing input complexity, indicating that local features of an instance play a major role in how hard it is to solve it, rather than this being an intrinsic property of the generator.
The only generator that produces ``easier'' instances than the others is Spike.
These are instances tailored for showing the difference between vertex
and point guards, requiring few guards to be placed in the middle of
certain free areas. We include the Spike instances in our experiments because they are an established class of test cases, being aware that all of the current implementations are able to identify good non-vertex positions for guards, and that this class has to be considered easy.

%% file: cps-agplib-description.tex
For the performed experiments, several classes of polygons were considered.
The majority of them were collected from AGPLib~\cite{art-gallery-instances-page}, which
is a library of sample instances for the \agp, consisting of various classes of
polygons of multiple sizes.
They include the test sets from many
previously published papers~\cite{csr-eeaoagp-07,csr-eeeaoagp-08,crs-exmvg-11,trs-qosagp-13,DaviPedroCid-J-002013,bfks-esbgagp-10,kbfs-esbgagp-12}.

To find out more about how each of the classes was generated,
see~\cite{csr-eeaoagp-07} and \cite{DaviPedroCid-J-002013}.
Below, we show a short description of the six classes of instances considered in
this survey; all of them are randomly generated:

%% file: tab-speedup-vguard.tex
\begin{table}
  \small
  \centering
  \begin{tabular}{|p{2.6cm}|c|r|r|r|r|r|r|}
    \hline
    \multicolumn{1}{|c|}{\multirow{2}{*}{\textbf{Class}}}
    & \multicolumn{1}{c|}{\multirow{2}{*}{\textbf{$n$}}}
    & \multicolumn{6}{c|}{\textbf{Speedup Factor}}\\
    &
    & \multicolumn{1}{c|}{\cix}
    & \multicolumn{1}{c|}{\bsx}
    & \multicolumn{1}{c|}{\cxiiisea}
    & \multicolumn{1}{c|}{\cxiiifull}
    & \multicolumn{1}{c|}{\bscur}
    & \multicolumn{1}{c|}{\ccur}
    \\ \hline
    \multirow{5}{*}{\vbox{Simple}}
    &200&1.00&0.66&1.03&1.21&7.54&\textbf{6.75}\\
    &500&1.00&0.66&1.01&1.02&7.79&\textbf{10.21}\\
    &1000&1.00&0.66&1.02&0.95&8.03&\textbf{14.65}\\
    &2000&1.00&0.68&1.00&0.90&10.24&\textbf{18.97}\\
    &5000&--&--&--&--&--&--\\
    \hline
    \multirow{5}{*}{\vbox{Orthogonal}}
    &200&1.00&0.64&1.01&1.05&\textbf{6.46}&6.15\\
    &500&1.00&0.63&1.01&0.98&6.67&\textbf{10.82}\\
    &1000&1.00&0.65&1.00&0.92&7.75&\textbf{15.67}\\
    &2000&1.00&0.65&0.98&0.82&9.57&\textbf{19.52}\\
    &5000&1.00&0.72&1.00&0.75&12.63&\textbf{28.64}\\
    \hline
     \multirow{5}{*}{\vbox{von Koch}}
     &200&1.00&0.38&1.02&1.33&2.09&\textbf{3.45}\\
     &500&1.00&0.44&1.09&1.37&1.86&\textbf{4.27}\\
     &1000&1.00&0.60&0.95&1.40&1.95&\textbf{4.75}\\
     &2000&1.00&0.92&1.34&1.39&2.67&\textbf{6.18}\\
     &5000&--&--&--&--&--&--\\
     \hline
   \end{tabular}
   \caption{
     Speedup for vertex guards. Numbers indicate how many times faster
     than \cix\ later implementations became, computed as log-average.
     The comparison is only possible when there is at least one instance
     of the group that was solved by all considered solvers.
 	This table is restricted to simple polygons, since \cix does not support polygons with holes.}
   \label{tab:speedup-vguard}
\end{table}

%% file: tab-gopt-frac.tex
\begin{table}\centering
  \begin{tabular}{|ll|rrrrr|rrrrr|}\hline
    &&\multicolumn{5}{c|}{\cidC{\bf polygons without holes}}        & \multicolumn{5}{c|}{\bf polygons with holes}  \\
    &&\BX{200}&\BX{500}&\BX{1000}&\BX{2000}&\BX{5000}&\BX{200}&\BX{500}&\BX{1000}&\BX{2000}&\BX{5000}\\\hline
    &\bsx&
    55.6 & 27.8 & 14.4 & 3.3 & 0.0 & 54.4 & 32.2 & 28.9 & 30.0 & 0.0
    \\
    OPT\hspace{1em}\mbox{}&\bsxii&
    53.3 & 30.0 & 11.1 & 4.4 & 0.0 & 54.4 & 32.2 & 27.8 & 31.1 & 0.0
    \\
    &\bscur&
    56.7 & 24.4 & 12.2 & 1.1 & 0.0 & 50.0 & 31.1 & 27.8 & 31.1 & 33.3
    \\\hline
    &\bsx&
    93.3&100.0&100.0&100.0&33.3&96.7&100.0&98.9&75.6&0.0
    \\
    5\% gap\hspace{1em}\mbox{}&\bsxii&
    93.3&100.0&100.0&100.0&33.3&97.8&98.9&98.9&72.2&0.0
    \\
    &\bscur&
    91.1&100.0&100.0&100.0&98.9&97.8&98.9&98.9&98.9&33.3
    \\\hline
  \end{tabular}
  \caption{Optimality rates for fractional point guards.}
  \label{tab:gopt-frac}
\end{table}

%% file: tab-opt-agp.tex
\begin{table}
  \small
  \centering
  \begin{tabular}{|p{2.6cm}|c|r|r|r|r|r|}
    \hline
    \multicolumn{1}{|c|}{\multirow{2}{*}{\textbf{Class}}}
    & \multicolumn{1}{c|}{\multirow{2}{*}{\textbf{$n$}}}
    & \multicolumn{5}{c|}{\textbf{Optimality Rate \dctC{(\%)}}} \\
    &
    & \multicolumn{1}{c|}{\bsxii}
    & \multicolumn{1}{c|}{\cxiiisea}
    & \multicolumn{1}{c|}{\cxiiifull}
    & \multicolumn{1}{c|}{\bscur}
    & \multicolumn{1}{c|}{\ccur}
    \\ \hline

    \multirow{5}{*}{\vbox{Simple}}
    & 200    &   \textbf{100.0}  &   \textbf{100.0}  &   \textbf{100.0}  &   96.7   &   \textbf{100.0}  \\
    & 500    &   76.7   &   \textbf{100.0}  &   \textbf{100.0}  &   96.7   &   \textbf{100.0}  \\
    & 1000   &   70.0   &   96.7   &   \textbf{100.0}  &   90.0   &   \textbf{100.0}  \\
    & 2000   &   36.7   &   6.7    &   50.0   &   60.0   &   \textbf{100.0}  \\
    & 5000   &   0.0    &   0.0    &   0.0    &   26.7   &   \textbf{100.0}  \\ \hline

    \multirow{5}{*}{\vbox{Orthogonal}}
    & 200     &   96.7   &   \textbf{100.0}  &   \textbf{100.0}  &   96.7   &   96.7   \\
    & 500     &   86.7   &   \textbf{100.0}  &   96.7   &   93.3   &   93.3   \\
    & 1000    &   70.0   &   \textbf{100.0}  &   \textbf{100.0}  &   86.7   &   \textbf{100.0}  \\
    & 2000    &   46.7   &   70.0   &   90.0   &   70.0   &   \textbf{100.0}  \\
    & 5000    &   0.0    &   0.0    &   0.0    &   40.0   &   93.3   \\ \hline \hline

    \multirow{5}{*}{\vbox{Simple-simple}}
    & 200     &   93.3  &  --   &   \textbf{100.0}  &   86.7   &   \textbf{100.0}  \\
    & 500     &   76.7  &  --   &   83.3   &   60.0   &   \textbf{100.0}  \\
    & 1000    &   3.3   &  --     &   0.0    &   13.3   &   \textbf{100.0}  \\
    & 2000    &   0.0   &  --    &   0.0    &   0.0    &   46.7   \\
    & 5000    &   0.0   &  --   &   0.0    &   0.0    &   0.0    \\ \hline

    \multirow{5}{*}{\vbox{Ortho-ortho}}
    & 200     &   83.3   & --   &   96.7   &   86.7   &   \textbf{100.0}  \\
    & 500     &   53.3   &  --    &   83.3   &   53.3   &   \textbf{100.0}  \\
    & 1000    &   16.7   & --      &   3.3    &   16.7   &   96.7   \\
    & 2000    &   0.0    & --     &   0.0    &   0.0    &   33.3   \\
    & 5000    &   0.0    & --     &   0.0    &   0.0    &   0.0    \\ \hline \hline

    \multirow{5}{*}{\vbox{von Koch}}
    & 200     &   \textbf{100.0}  &   \textbf{100.0}  &   \textbf{100.0}  &   \textbf{100.0}  &   \textbf{100.0}  \\
    & 500     &   \textbf{100.0}  &   96.7   &   \textbf{100.0}  &   93.3   &   \textbf{100.0}  \\
    & 1000    &   \textbf{100.0}  &   46.7   &   \textbf{100.0}  &   96.7   &   \textbf{100.0}  \\
    & 2000    &   83.3   &   0.0    &   0.0    &   86.7   &   \textbf{100.0}  \\
    & 5000    &   0.0    &   0.0    &   0.0    &   0.0    &   0.0    \\ \hline

    \multirow{5}{*}{\vbox{Spike}}
    & 200     &   \textbf{100.0} &  --   &   \textbf{100.0}  &   96.7   &   \textbf{100.0}  \\
    & 500     &   \textbf{100.0} & --    &   \textbf{100.0}  &   \textbf{100.0}  &   \textbf{100.0}  \\
    & 1000    &   3.3 & --      &   96.7   &   \textbf{100.0}  &   \textbf{100.0}  \\
    & 2000    &   0.0 & --      &   96.7   &   \textbf{100.0}  &   \textbf{100.0}  \\
    & 5000    &   0.0 & --      &   0.0    &   96.7   &   \textbf{100.0}  \\
    \hline
  \end{tabular}
  \caption{Optimality Rate for point guards.}
  \label{tab:opt-agp}
\end{table}

%% file: tab-opt-no-proof-agp.tex
\begin{table}
  \small
  \centering
  \begin{tabular}{|p{2.6cm}|c|r|r|r|r|r|rrrrrrrrrrrrrrrrr}
    \hline
    \multicolumn{1}{|c|}{\multirow{2}{*}{\textbf{Class}}}
    & \multicolumn{1}{c|}{\multirow{2}{*}{\textbf{$n$}}}
    & \multicolumn{5}{c|}{\textbf{Optimality Rate \dctC{(\%)} without proof}} \\
    &
    & \multicolumn{1}{c|}{\bsxii}
    & \multicolumn{1}{c|}{\cxiiisea}
    & \multicolumn{1}{c|}{\cxiiifull}
    & \multicolumn{1}{c|}{\bscur}
    & \multicolumn{1}{c|}{\ccur}
    \\ \hline

    \multirow{5}{*}{\vbox{Simple}}
    &200& 	\textbf{100.0}&	\textbf{100.0}&	\textbf{100.0}&	96.7&	\textbf{100.0}\\
    &500&	80.0&	\textbf{100.0}&	\textbf{100.0}&	\textbf{100.0}&	\textbf{100.0}\\
    &1000&	73.3&	\textbf{100.0}&	\textbf{100.0}&	\textbf{100.0}&	\textbf{100.0}\\
    &2000&	50.0&	50.0&	80.0&	93.3&	\textbf{100.0}\\
    &5000&	0.0&	0.0&	0.0&	83.3&	\textbf{100.0}\\
    \hline
    \multirow{5}{*}{\vbox{Orthogonal}}
    &200&	96.7&	\textbf{100.0}&	\textbf{100.0}&	96.7&	96.7\\
    &500&	86.7&	\textbf{100.0}&	\textbf{100.0}&	93.3&	93.3\\
    &1000&	70.0&	\textbf{100.0}&	\textbf{100.0}&	90.0&	\textbf{100.0}\\
    &2000&	50.0&	96.7&	93.3&	90.0&	\textbf{100.0}\\
    &5000&	0.0&	0.0&	0.0&	50.0&	93.3\\
    \hline
    \hline
    \multirow{5}{*}{\vbox{Simple-simple}}
    &200&	96.7&	--&	\textbf{100.0}&	90.0&	\textbf{100.0}\\
    &500&	93.3&	--&	96.7&	80.0&	\textbf{100.0}\\
    &1000&	33.3&	--&	20.0&	73.3&	\textbf{100.0}\\
    &2000&	0.0&	--&	0.0&	33.3&	50.0\\
    &5000&	0.0&	--&	0.0&	0.0&	0.0\\
    \hline
    \multirow{5}{*}{\vbox{Ortho-ortho}}
    &200&	93.3&	--&	\textbf{100.0}&	\textbf{100.0}&	\textbf{100.0}\\
    &500&	80.0&	--&	93.3&	90.0&	\textbf{100.0}\\
    &1000&	70.0&	--&	30.0&	70.0&	96.7\\
    &2000&	0.0&	--&	0.0&	30.0&	43.3\\
    &5000&	0.0&	--&	0.0&	0.0&	0.0\\
    \hline
    \hline
    \multirow{5}{*}{\vbox{von Koch}}
    &200&	\textbf{100.0}&	\textbf{100.0}&	\textbf{100.0}&	\textbf{100.0}&	\textbf{100.0}\\
    &500&	\textbf{100.0}&	\textbf{100.0}&	\textbf{100.0}&	93.3&	\textbf{100.0}\\
    &1000&	\textbf{100.0}&	70.0&	\textbf{100.0}&	96.7&	\textbf{100.0}\\
    &2000&	83.3&	0.0&	30.0&	90.0&	\textbf{100.0}\\
    &5000&	0.0&	0.0&	0.0&	0.0&	0.0\\
    \hline
    \multirow{5}{*}{\vbox{Spike}}
    &200&	\textbf{100.0}&	--&	\textbf{100.0}&	96.7&	\textbf{100.0}\\
    &500&	\textbf{100.0}&	--&	\textbf{100.0}&	\textbf{100.0}&	\textbf{100.0}\\
    &1000&	3.3&	--&	\textbf{100.0}&	\textbf{100.0}&	\textbf{100.0}\\
    &2000&	0.0&	--&	96.7&	\textbf{100.0}&	\textbf{100.0}\\
    &5000&	0.0&	--&	0.0&	96.7&	\textbf{100.0}\\
    \hline
  \end{tabular}
  \caption{Optimality Rate without proof for point guards.}
  \label{tab:opt-with-no-proof-agp}
\end{table}

%% file: tab-gap5-agp.tex
\begin{table}
  \small
  \centering
  \begin{tabular}{|p{2.6cm}|c|r|r|r|r|r|rrrrrrrrr}
    \hline
    \multicolumn{1}{|c|}{\multirow{2}{*}{\textbf{Class}}}
    & \multicolumn{1}{c|}{\multirow{2}{*}{\textbf{$n$}}}
    & \multicolumn{5}{c|}{\textbf{5\% gap Rate in \dctC{(\%)}}} \\
    &
    & \multicolumn{1}{c|}{\bsxii}
    & \multicolumn{1}{c|}{\cxiiisea}
    & \multicolumn{1}{c|}{\cxiiifull}
    & \multicolumn{1}{c|}{\bscur}
    & \multicolumn{1}{c|}{\ccur}
    \\ \hline

    \multirow{5}{*}{\vbox{Simple}}
    &200& 	100.0&	100.0&	100.0&	100.0&	100.0\\
    &500&	100.0&	100.0&	100.0&	100.0&	100.0\\
    &1000&	100.0&	100.0&	100.0&	100.0&	100.0\\
    &2000&	100.0&	100.0&	96.7&	100.0&	100.0\\
    &5000&	0.0&	0.0&	0.0&	100.0&	100.0\\
    \hline
    \multirow{5}{*}{\vbox{Orthogonal}}
    &200&	100.0&	100.0&	100.0&	100.0&	100.0\\
    &500&	100.0&	100.0&	100.0&	100.0&	100.0\\
    &1000&	100.0&	100.0&	100.0&	100.0&	100.0\\
    &2000&	100.0&	100.0&	100.0&	100.0&	100.0\\
    &5000&	0.0&	0.0&	0.0&	100.0&	100.0\\
    \hline
    \hline
    \multirow{5}{*}{\vbox{Simple-simple}}
    &200&	100.0&	--&	100.0&	100.0&	100.0\\
    &500&	100.0&	--&	93.3&	100.0&	100.0\\
    &1000&	100.0&	--&	33.3&	100.0&	100.0\\
    &2000&	0.0&	--&	0.0&	96.7&	80.0\\
    &5000&	0.0&	--&	0.0&	0.0&	0.0\\
    \hline
    \multirow{5}{*}{\vbox{Ortho-ortho}}
    &200&	100.0&	--&	100.0&	100.0&	100.0\\
    &500&	100.0&	--&	100.0&	100.0&	100.0\\
    &1000&	100.0&	--&	40.0&	96.7&	100.0\\
    &2000&	56.7&	--&	0.0&	76.7&	86.7\\
    &5000&	0.0&	--&	0.0&	0.0&	0.0\\
    \hline
    \hline
    \multirow{5}{*}{\vbox{von Koch}}
    &200&	100.0&	100.0&	100.0&	100.0&	100.0\\
    &500&	100.0&	100.0&	100.0&	100.0&	100.0\\
    &1000&	100.0&	73.3&	100.0&	100.0&	100.0\\
    &2000&	100.0&	0.0&	56.7&	100.0&	100.0\\
    &5000&	0.0&	0.0&	0.0&	3.3&	0.0\\
    \hline
    \multirow{5}{*}{\vbox{Spike}}
    &200&	100.0&	--&	100.0&	96.7&	100.0\\
    &500&	100.0&	--&	100.0&	100.0&	100.0\\
    &1000&	3.3&	--&	96.7&	100.0&	100.0\\
    &2000&	0.0&	--&	96.7&	100.0&	100.0\\
    &5000&	0.0&	--&	0.0&	96.7&	100.0\\
    \hline
  \end{tabular}
  \caption{Rate of upper bound within 5\% distance to lower bound.}
  \label{tab:opt-with-gap}
\end{table}

%% file: speedups.tex
\section{Success Factors}
\label{sec:speedup}

\cidC{As  seen  in  Section~\ref{sec:time},  the  most  effective
  algorithms for the \agp can be decomposed into four elements:}\todo{R2.G1}
\begin{itemize}
\item Geometric subroutines dealing with computing visibility
  relations, determining feasibility,
\item Set Cover subroutines computing (near-)optimal solutions for
  finite cases,
\item Routines to find candidates for discrete guard and witness
  locations, and
\item An outer algorithm combining the three parts above.
\end{itemize}
In this section, we focus on these techniques.

\subsection{Geometric Subroutines}\label{sec:speedup.geometry}

Both groups use the 2D Arrangements package~\cite{cgal:wfzh-a2-12}
of \cgal which follows
the {\em generic programming paradigm}~\cite{a-gps-99}. For instance,
in the case of arrangements it is possible to change the curve type
that is used to represent the planar subdivisions or the kernel that
provides the essential geometric operations and also determines the
number type used. In the context of this work, it is clear that the
used curves are simply segments%
\footnote{In the context of fading~\cite{ks-eagi-12}
circular arcs may also be required.}.
However, the choice of the geometric kernel can have a significant
impact on the runtime.

First of all, it should be noted that among the different kernels that
\cgal offers only kernels that provide exact constructions should be
considered as any inexact construction is likely to induce
inconsistencies in the data structure of the arrangements package.
This already holds for seemingly simple scenarios as the code
of the arrangement package heavily relies on the assumption that
constructions are exact.

This essentially leaves two kernels:
The Cartesian kernel and the lazy-exact kernel. For both kernels
it is possible to exchange the underlying exact rational number type,
but {\tt CGAL::Gmpq}~\cite{gmp} is the recommended one%
\footnote{Other options are, for instance,
{\tt leda::rational}~\cite{mn-lpcgc-00} or
{\tt CORE::BigRat}~\cite{core},
but, compared to Gmpq, both imply some overhead and are only recommended
in case the usage of the more complex number types of these libraries
is required.}.

The {\bf Cartesian kernel}, is essentially the naive
application of exact rational arithmetic
(using the one that it is instantiated with, in this case {\tt CGAL::Gmpq}). \
Thus, coordinates are represented by
a numerator and denominator each being an integer using as many
bits as required.
This implies that even basic geometric constructions and predicates
are not of constant cost, but depend on the bit-size of their input.
For instance, the intersection point of two segments is likely to
require significantly more bits than the endpoints of the segments.
And this is even more relevant in case of cascaded constructions
as the bit growth is cumulative.
This effect is very relevant in both approaches due to there iterative
nature, e.g., when such a point is chosen to be a new guard or
witness position.

The {\bf lazy-exact kernel}~\cite{pf-glese-06} tries to attenuate all these
effects
by using exact arithmetic only when necessary.
Every arithmetic operation and construction is first carried out
using only double interval arithmetic, that is, using directed rounding,
an upper and a lower of the exact value is computed.
The hope is that for most cases this is already sufficient to give
the correct and certified answer, for instance whether a point is
above or below a line.
However, for the case when this is not sufficient, each constructed
object also knows its history, which makes it possible to carry out
the exact rational arithmetic as it is done in the Cartesian kernel
in order to determine the correct result. This idea is implemented by
the lazy kernel not only on the number type level%
\footnote{This can be achieved by the instantiation of the Cartesian kernel
with {\tt CGAL::Lazy\_exact\_nt<CGAL::Gmpq>}.}, but also for predicates and
constructions, which reduces the overhead (memory and time) that is induced
by maintaining the history.

\begin{table}
\begin{center}
\begin{tabular}{|l|ccccc|}
  \hline
  Class \ Size& 200&	500&	1000&	2000&	5000\\
  \hline
  Simple&	1.27&	1.46&	1.55&	1.49&	1.35\\
  Orthogonal&	1.44&	1.60&	1.66&	1.69&	1.65\\
  Simple-simple&2.15&	1.72&	1.44&	1.37&	-\\
  Ortho-ortho&	1.54&   1.30&	1.21&	1.20&	-\\
  von~Koch&	1.02&	1.06&	1.10&	1.16&	-\\
  Spike&	1.15&	1.61&	1.76&	2.10&	2.56\\
  \hline
\end{tabular}
\end{center}
\caption{
  \label{tab:camp-cart-vs-epec}%
  The speedup factor of {\ccur} using the Cartesian kernel and
  the lazy-exact kernel. Similar numbers were obtained for \bsxii.
  The lazy-exact kernel is now the standard configuration in \bscur and \ccur.
}
\end{table}

By the genericity of \cgal it is possible to easily exchange
the used  geometric kernel.
Table~\ref{tab:camp-cart-vs-epec} shows the speedup factors
by using the Cartesian kernel vs the lazy-exact kernel for
the different instances for \cxiiisea. It should be noted that all
Braunschweig and Campinas implementations since 2007 use a complexity
reduction step together with the Cartesian kernel: Whenever a point in
a face is generated, it is rounded to a nearby point of lower bit
complexity. Without this, neither implementation would be able to
solve any instance of substantial size. This speedup technique is
missing in the variant with the lazy-exact kernel, as it requires
to actually compute the point coordinates before rounding, which would
defeat the purpose of the kernel. Therefore the table compare the
lazy-exact kernel against the Cartesian kernel with explicit
complexity reduction.

For the random polygons, as well as for the spike ones, it can be
observed that the lazy-exact kernel is usually almost twice as
fast as the Cartesian kernel.
However, for the von Koch polygons the lazy-exact
kernel only gives a mild speedup. We explain this by two effects.
First, the bit-size of the input polygons is not very large and also
the bit-size of intermediate constructions do not grow as much,
as the horizontal and vertical lines dominate the scene.
Second, the instance induces degenerate
situations in which the lazy-exact kernel must fall back to the exact
arithmetic in which cases effort for interval arithmetic and
maintaining the history is a real overhead.
The lazy-exact kernel is now the standard
configuration in \bscur and \ccur.

\subsubsection{Visibility Computation}

\begin{figure}
  \centering
  \begin{subfigure}[b]{0.48\textwidth}
    \includegraphics[width=\textwidth]{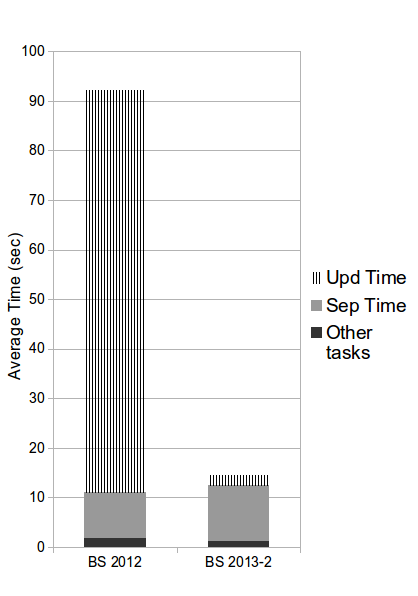}
    \caption{Braunschweig}
  \end{subfigure}
  ~
  \begin{subfigure}[b]{0.48\textwidth}
    \includegraphics[width=\textwidth]{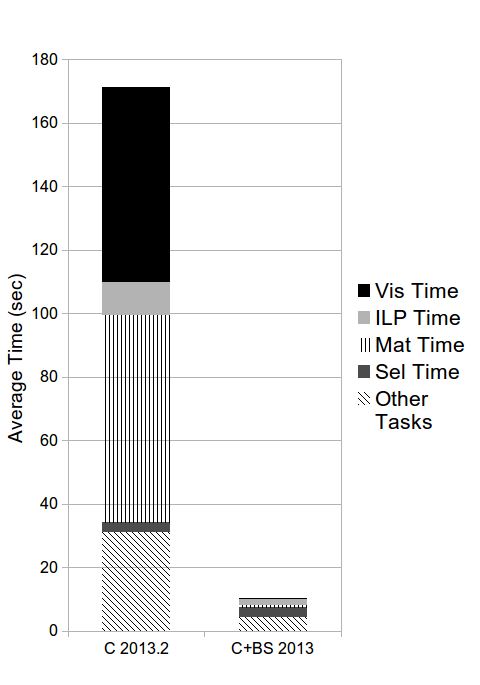}
    \caption{Campinas}
  \end{subfigure}
  \caption{
    \label{fig:visbility_bars}
    Split up of average total time for different configurations on
    all simple instances with 1000 vertices.
    (left)
    The update time which is dominated by the visibility
    polygon computation almost vanishes in \bscur
    compared to the \bsxii.
    (right)
    The time spent on visibility in \ccur
    is almost negligible compared to the time spend in \cxiiifull.
   }
\end{figure}

One of the most significant improvements with respect to speed is due to the
new upcoming visibility package~\cite{cgal:visibility} of \cgal.
This package was developed by the group in Braunschweig with this
project being the main motivation.
Of course, this packages was also made available to the
group in Campinas prior to its actual integration in \cgal.
Figure~\ref{fig:visbility_bars} illustrates the tremendous impact
on the runtime for both approaches.
The left side shows the split up of total runtime for the code from Braunschweig
in 2012 and 2013. While in 2012 the update time (dominated by visibility computation)
used about two third of the time for visibility computation is now almost
negligible. The same holds for improvements achieved in the code from Campinas,
see right side of Figure~\ref{fig:visbility_bars}.
It can  be noticed in  the latter graph  that the time  spent by
\ccur\ in building the  constraint matrices for the {\ilp}s, denoted
by  {\tt Mat  Time},  also  suffered a  huge  reduction relative  to
\cxiiifull. As commented in Section~\ref{campinas2013-cur}, this was
mostly  due to  the execution  of  the visibility  testing from  the
perspective of the witnesses rather than the guards.

\subsection{Set Cover Optimization}\label{sec:speedup.sc}

Many \agp algorithms rely on repeatedly solving $\AGP(G,W)$
(Equations~\eqref{eq.infip.obj}--\eqref{eq.infip.bin}) for finite $G$ and $W$
as a subroutine, corresponding to the NP-hard \scp.
Therefore improving the solutions times for these \scp
instances can benefit the overall algorithm.

\subsubsection{Lagrangian Relaxation}\label{sec:lagrange}

\input{speedups-scp-campinas.tex}

\subsubsection{DC Programming} A different solution method for the
Braunschweig approach was discussed in Kr{\"o}ller
\etal~\cite{kms-neaagp-13}. Here, the \ilp representing the \scp for $\AGP(G,W)$ was
rewritten as
\begin{equation}
  \min_{x\in\dsR^G} F(x)\;\text{, where}\; F(x) := \sum_{g\in G}x_g - \theta
  \sum_{g\in G} x_g(x_g-1) + \chi(x) \;.
\end{equation}
Here, $\theta$ is a sufficiently large constant used to penalize
fractional values for $x_g$, and $\chi\colon \dsR^G\to \{0,\infty\}$ is an
indicator function with
\begin{equation}
  \chi(x) = 0 \;:\!\iff\; \left\{
    \begin{array}{ll}
      \sum_{g\in\vis{w}} x_g \geq 1 &\forall w\in W\\
      0\leq x_g\leq 1 & \forall g\in G
    \end{array}
    \right.
    \;.
\end{equation}
It is easy to see that $F$ can be expressed as $F(x):=f_1(x)-f_2(x)$,
where
\begin{equation}
  f_1(x):=\sum_{g\in G} x_G + \chi(x),
  \quad\text{and}\quad
  f_2(x)=\theta\sum_{g\in G} x_g(x_g-1)\;,
\end{equation}
i.e., the \scp instance is reduced to minimizing the difference
of two non-linear convex functions. For such optimization problems,
the DCA algorithm~\cite{pl-dca-97} can be used. In experiments, it was
shown that solutions for $\AGP(G,W)$ could be found very quickly,
however, at the time, the large runtime overhead of the geo\-metric
subroutines led to inconclusive results on the benefits. Revisiting
this approach with the new \bscur\ and \ccur\ implementations, which
no longer suffer from this overhead, will be an interesting experiment
left for future work.

\subsection{Point Generation}\label{sec:speedup.pointgen}

The central heuristic component in most \agp implementations are point
generators, which choose where to place new guards and witnesses. One
cannot expect these problems to be simple, given that a perfect
choice for $G$ and $W$ equals solving \agp optimally.

\subsubsection{Guard Placement} One subroutine in the algorithms is
to improve the current set of guards, given a current set $W$ of
witnesses. This corresponds to finding guards that can be used to
solve $\AGP(P,W)$.

A critical observation~\cite{trs-qosagp-13} allows for elegant
solution to this problem: Consider the visibility arrangement
$\visarr(W)$. It is always possible to find an optimal solution for
$\AGP(P,W)$ where each \avp contains at most one guard. This can be
strengthened by observing that the guards can be restricted further to
light \avps. As explained in Section~\ref{campinas2013}, the \cxiiisea\
algorithm uses as guard candidates the vertices of $P$ along with all
vertices from light \avps. In \ccur, a second guard placement strategy
using no more than one interior point per \avp is available.
Results comparing these two can  be  seen in
Table~\ref{tab:guards-campinas}. It  is possible to conclude that
the  latest  guard  placement  strategy,  which  consists  of  using
only one point within each light \avp, is often the best option.
The explanation for this success is probably related to the fact that, with
the winning strategy, there is a reduced number of visibility tests between witnesses
and guard candidates, as well as a smaller size of \scp instances to be solved.
\begin{table}
  \begin{center}
    \begin{tabular}{|r|c|c|}
      \hline
      & Vertices of light \avps& Interior of light \avps \\
      \hline
      simple              2000 & 100.0 & 100.0 \\
      ortho               2000 & 100.0 & 100.0 \\
      simple-simple       2000 &   6.7 & 33.3 \\
      ortho-ortho         2000 &  13.3 & 46.7 \\
      von Koch            2000 & 100.0 & 100.0 \\
      spike               2000 & 100.0 & 100.0 \\
      \hline
    \end{tabular}
  \end{center}
  \caption{Percentage of instances solved to binary optimality by
    the current implementation from Campinas with guard candidates
    on vertices or inside light \avps.}
  \label{tab:guards-campinas}
\end{table}

For $\AGPFrac(P,W)$, as solved by the Braunschweig line of algorithms,
this observation can be extended further: If an optimal dual solution
for $\AGPFrac(G,W)$ is available, selecting additional guards
corresponds to a column generation process. Therefore, the
BS algorithms place guards only in light \avps where the dual solution
guarantees an improvement in the objective function. To avoid cycling
in the column generation process, $G$ is monotonically growing, leading over time to
a large number of guard positions.

\subsubsection{Witness Placement} 
\begin{figure}\centering
  \def\svgwidth{.5\textwidth}
  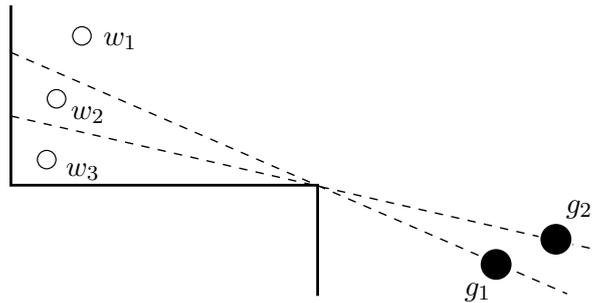
  \caption{Creeping shadow effect.}
  \label{fig:shadow}
\end{figure}

The choice of witnesses is as important
as that of the guards. In principle, the same reasoning as for guards can be used:
Given guard candidates $G$, creating $W$ with one witness in every shadow \avp of $\visarr(G)$
guarantees that a solution for $\AGP(G,W)$ is also a solution for
$\AGP(G,P)$. A na{\"i}ve placement algorithm based on this observation
would simply create witnesses in shadow \avps. However, this leads to the
problem of creeping shadows at reflex vertices, see
Figure~\ref{fig:shadow}: Placing witness in the interior of the \avp
adjacent to the polygon boundary creates an infinite chain of
guard/witness positions that converges towards a witness on the
boundary, but not reaching it. Both the Braunschweig and the Campinas
algorithms therefore can create additional witnesses on the edges of
shadow \avps.

\subsubsection{Initial Set} The selection of the first candidates for
guards and witnesses, i.e., the initial choice of $G$ and $W$ can have
tremendous impact on algorithm runtime. In principle, a good heuristic
here could pick an almost optimal set for $G$ and a matching $W$ to
prove it, and reduce the algorithm afterwards to a few or even no
iterations.

Chwa~\etal~\cite{cbkmos-gaggw-06} provide a partial answer to this
problem: They attempt to find a finite set of witnesses with the
property that guarding this set guarantees guarding the whole
polygon. If such a set exists, the polygon is called {\em
  witnessable}. Unfortunately this is not always the case. However,
for a witnessable polygon, the set can be characterized and quickly
computed. Current algorithms do not bother checking for witnessability
(although Chwa \etal~provide an algorithm), but rather directly
compute this set and use it for initial witnesses.
Should the polygon be witnessable, the algorithm automatically terminates
in the first iteration.

Considering the current version from Campinas, two initial discretizations are
used: \cv  and  \cp.
The first one includes only the convex vertices of the polygon in the
initial   set, while  the   second  chooses   the  middle   points  of
reflex-reflex  edges and  the  convex vertices  that  are adjacent  to
reflex vertices.

\begin{figure}
  \centering
  \includegraphics[width=0.96\textwidth]{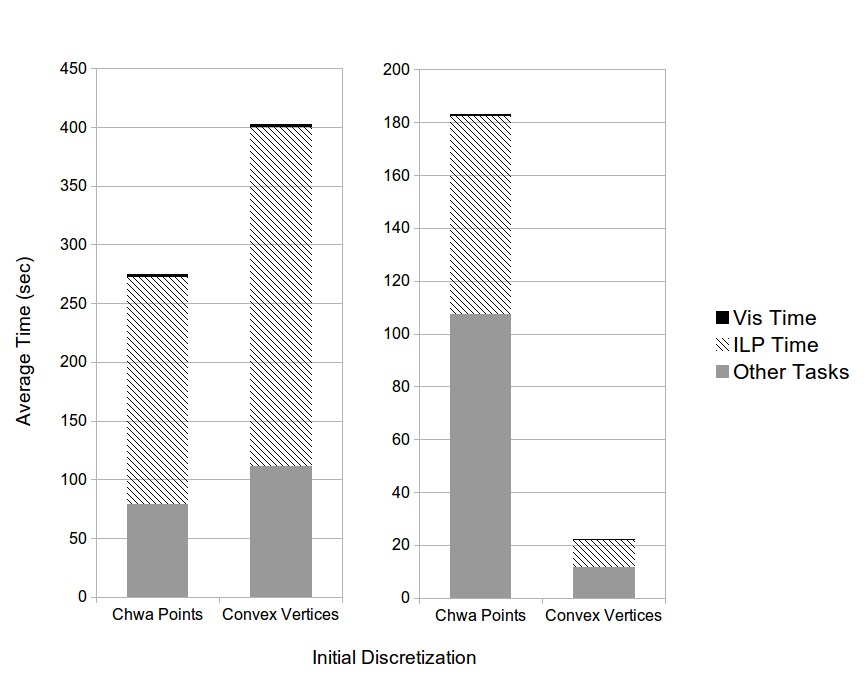}

  \caption{Average Time needed to solve ortho-ortho (left) and spike (right)
      polygons with 1000 vertices using the \acl{CV} and the
    \acl{CP} discretization.}
  \label{fig:chwa-convex}
\end{figure}

The two charts in Figure~\ref{fig:chwa-convex}
show the average run time necessary to
find optimal solutions when using \cv and \cp strategies on
simple-simple and spike polygons.
From these charts, one can perceive that there is an
advantage in using the
\cp discretization for polygons from the simple-simple class.
On the other hand, the chart corresponding to the
spike polygons shows that the implementation works much better
when the strategy chosen is the \cv one.
In this last case, the program required four times less time to solve
the same set of  polygons when using the \cv strategy as opposed to \cp.

For \bsx, several strategies were implemented, see Table~\ref{tab:speedup.bsGW} which is extracted from the corresponding paper~\cite{kbfs-esbgagp-12}:
Leaving $G$ and $W$ empty (for implementation reasons, both contained one arbitrary point),
putting guards and witnesses on every (or every other) vertex of the polygon,
putting guards on all reflex vertices,
and putting a witness on every edge adjacent to a reflex vertex.
The Chwa-inspired combination allowed for a speedup of around two.

\begin{table}
  \centering
  \begin{tabular}{|ll|r|}\hline
    Initial $G$        & Initial $W$        & Speedup \\\hline
    Single Point       & Single Point       &    1.00 \\
    Every other vertex\hspace{1em}\mbox{} & Every other vertex &    1.59 \\
    All vertices       & All vertices       &    1.64 \\
    All vertices       & Reflex edges       &    1.74 \\
    Reflex vertices    & Reflex edges       &    2.02 \\\hline
  \end{tabular}
  \caption{Speedup factors in \bsx\ obtained by varying initial guards
    and witnesses~\cite{kbfs-esbgagp-12}.}
  \label{tab:speedup.bsGW}
\end{table}

\subsection{Lower Bounds}
\label{sec:speedup.lb}

A crucial success factor for solving the binary \agp variants is the
quality of the lower bounds.  This is especially visible in \bscur, which was tested with and without the
cutting planes, \ie, with and without the features published
in~\cite{f-isagplp-12,ffks-ffagp-14} and outlined in Section~\ref{sec:bs2012}.
Table~\ref{tab:cuts} compares the solution rates for the different
classes of instances with 500 vertices and clearly shows that using
cutting planes greatly improves solution rates.  Cutting planes
increase the lower bounds and improve the solution rates for all classes
of instances.
\begin{table}
	\begin{center}
	\begin{tabular}{|r|c|c|}
		\hline
		Class / Technique & With Cutting Planes & Without Cutting Planes \\
		\hline
		ortho         &80.0\%  &63.3\% \\
		simple        &86.7\%  &40.0\% \\
		von Koch      &100.0\% &70.0\% \\\hline
		ortho-ortho   &63.3\%  &13.3\% \\
		simple-simple &70.0\%  &6.7\% \\
		spike         &100.0\% &96.7\% \\
		\hline
	\end{tabular}
	\end{center}
	\caption{%
          Percentage of instances solved to binary optimality comparing
          two variants of code from Braunschweig 2013,
          one with and without cutting planes, for 500-vertex instances.}
	\label{tab:cuts}
\end{table}

\input{cps-lb-description}

%% file: speedups-scp-campinas.tex
In the  algorithm developed by  the research group in  Campinas
subsequent to the journal version from 2013~\cite{DaviPedroCid-J-002013}
(Section~\ref{campinas2013J}), {attempts were made} to reduce the
time spent by the \ilp solver through the implementation of some known
techniques, such as \ilp matrix reduction and Lagrangian heuristic.

{A standard method  for reducing constraints  and variables was used,
which}
is  based  on   {inclusion  properties}  among columns  (guard
candidates)  and  rows  (witnesses)  of the  Boolean  {constraint
matrix of the \ilp that models the \scp instance.}

\begin{figure} \centering
  \includegraphics[width=0.7\textwidth]{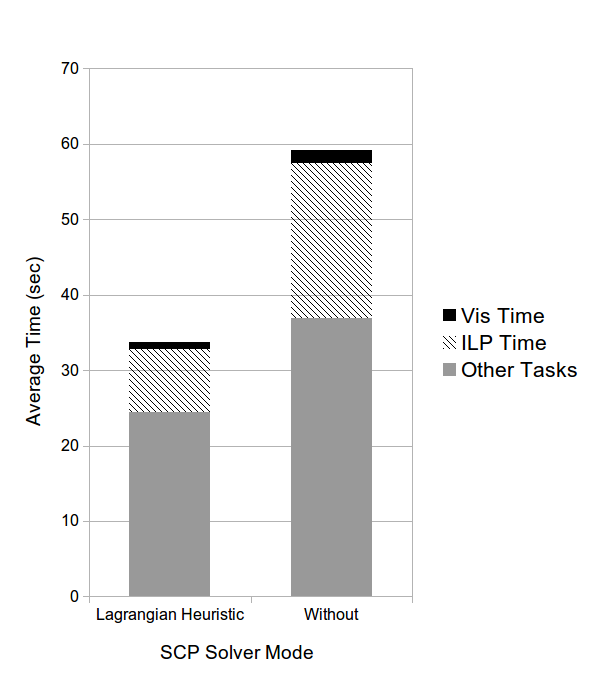}
  \caption{Average Time needed for the current Campinas version to solve
   von Koch polygons with 1000 vertices {with and without} the
    Lagrangian Heuristic.}
  \label{fig:lagran-comp}
\end{figure}

Furthermore, their algorithm {employs} a {\em  Lagrangian Heuristic} in  order
to obtain {good, hopefully  optimal, feasible starting solutions for the \scp to
speedup the convergence towards an optimum.} See
\cite{Beasley1993} for a comprehensive introduction to this technique.
The heuristic implemented is based on the work presented in~\cite{Beasley1993}.
Figure~\ref{fig:lagran-comp} shows how the use of this technique positively
influenced the average run time of the approach.

%% file: shadow2.pdf_tex
\begingroup%
  \makeatletter%
  \providecommand\color[2][]{%
    \errmessage{(Inkscape) Color is used for the text in Inkscape, but the package 'color.sty' is not loaded}%
    \renewcommand\color[2][]{}%
  }%
  \providecommand\transparent[1]{%
    \errmessage{(Inkscape) Transparency is used (non-zero) for the text in Inkscape, but the package 'transparent.sty' is not loaded}%
    \renewcommand\transparent[1]{}%
  }%
  \providecommand\rotatebox[2]{#2}%
  \ifx\svgwidth\undefined%
    \setlength{\unitlength}{173.56135645bp}%
    \ifx\svgscale\undefined%
      \relax%
    \else%
      \setlength{\unitlength}{\unitlength * \real{\svgscale}}%
    \fi%
  \else%
    \setlength{\unitlength}{\svgwidth}%
  \fi%
  \global\let\svgwidth\undefined%
  \global\let\svgscale\undefined%
  \makeatother%
  \begin{picture}(1,0.49419728)%
    \put(0,0){\includegraphics[width=\unitlength]{shadow2.pdf}}%
    \put(0.77401496,0.01115456){\color[rgb]{0,0,0}\makebox(0,0)[b]{\smash{$g_1$}}}%
    \put(0.94185738,0.14873803){\color[rgb]{0,0,0}\makebox(0,0)[b]{\smash{$g_2$}}}%
    \put(0.18291027,0.42810511){\color[rgb]{0,0,0}\makebox(0,0)[b]{\smash{$w_1$}}}%
    \put(0.12936489,0.304718){\color[rgb]{0,0,0}\makebox(0,0)[b]{\smash{$w_2$}}}%
    \put(0.12121672,0.21043159){\color[rgb]{0,0,0}\makebox(0,0)[b]{\smash{$w_3$}}}%
  \end{picture}%
\endgroup%

%% file: cps-lb-description.tex
For the Campinas approach, the quality of the lower bound computed
is a very important issue.
For $\AGP(P,P)$, the lower bound is obtained by solving an
$\AGP(P,W)$ instance, where $W$ is a discretized set of witnesses
points within $P$.
Therefore, it is fair to say that the quality of the value computed is
directly dependent on the strategy applied to select the points that
comprise the set $W$.
For more information on how the witness set is managed and how it affects
convergence of Campinas method, see
Section~\ref{sec:speedup.pointgen}.

%% file: open.tex
\section{Variants and Open Problems}
\label{sec:open}

\subsection{Fading}
\label{sec:open.fading}

An interesting variant for the \agp was proposed by Joe
O'Rourke in 2005: What if visibility suffers from fading effects, just
like light in the real world does? To be precise, we assume that for a
guard $g$ with intensity $x_g$, a witness $w\in\vis{g}$ is illuminated
with a value of $\varrho(d(g,w))x_g$, where $d(g,w)$ is the Euclidean
distance between $g$ and $w$, and $\varrho$ is a fading function,
usually assumed to be
\begin{equation}
  \varrho(d) := \left\{
    \begin{array}{cl}
      1 &\text{ if } d<1 \\
      d^{-\alpha} &\text{ if } 1\leq d< R\\
      0 & \text{ if } d\geq R
    \end{array}\right. \; .
\end{equation}
Here, $\alpha$ is a constant (2 for natural light in 3D space), and
$R$ is a maximal radius beyond which illumination is neglected. Fixing
$\varrho(d)$ to $1$ for small $d$ is necessary to keep the problem
well-defined. Otherwise, an infinitesimally small light can illuminate
a small circle around it. Then, no finite solution can exist, because
it can always be improved by creating additional guards between the
existing ones, and reducing intensity for all. This converges towards
the setup of $G=P$, with all $x_g=0$, which is not feasible.

Very little is known about this variant. A restricted case has been
discussed by Eisenbrand \etal\ \cite{efkm-esi-05}, where a
1-dimensional line segment is illuminated from a fixed set of guards. It
is shown how to solve this problem exactly and approximatively using
techniques from mathematical programming.

The primal-dual Braunschweig algorithm was shown to apply to this
variant as well: Kr{\"o}ller \etal\ \cite{ks-eagi-12} have modified the
\ilp formulation~\eqref{eq.infip.obj}--\eqref{eq.infip.bin} to use the
constraint
\begin{equation}
  \sum_{g\in\vis{w}} \varrho(d(g,w))x_g \geq 1 \;\; \forall w\in W
\end{equation}
instead of~\eqref{eq.infip.cover}. Two algorithms for vertex guards were
proposed and tested~\cite{EFHKKMS-AAGI-2014}, based on the
\bscur\ implementation. The first approximates $\varrho$ with a
step function, and uses updated primal and dual separation routines that
operate on overlays of visibility polygons and circular arcs,
resulting in an FPTAS for
the fractional $\AGP(V,P)$. The
other is based on continuous optimization techniques, namely a simplex
partitioning approach. In an experimental evaluation using polygons
with up to 700 vertices, it was found that most polygons can be solved
(to an 1.2-approximation in case of the discrete approach) within 20
minutes on a standard PC. The continuous algorithm turned out to be
much faster, and very often finishing with an almost-optimal solution
with a gap under 0.01\%. In an experimental work by Kokem{\"u}ller~\cite{k-vagp-14},
$\AGP(P,P)$ with fading was analyzed. It was found that placing guards
makes the problem substantially more difficult. This is mainly due to
an effect where moving one guard requires moving chains of other
guards as well to cover up for decreased illumination. It was also found that
scaling an input polygon has an impact on the structure of solutions
and number of required guards, resulting in a dramatic runtime impact.

\subsection{Degeneracies}
\label{sec:open.degen}

\begin{figure}
  \centering
  \includegraphics[width=.5\textwidth]{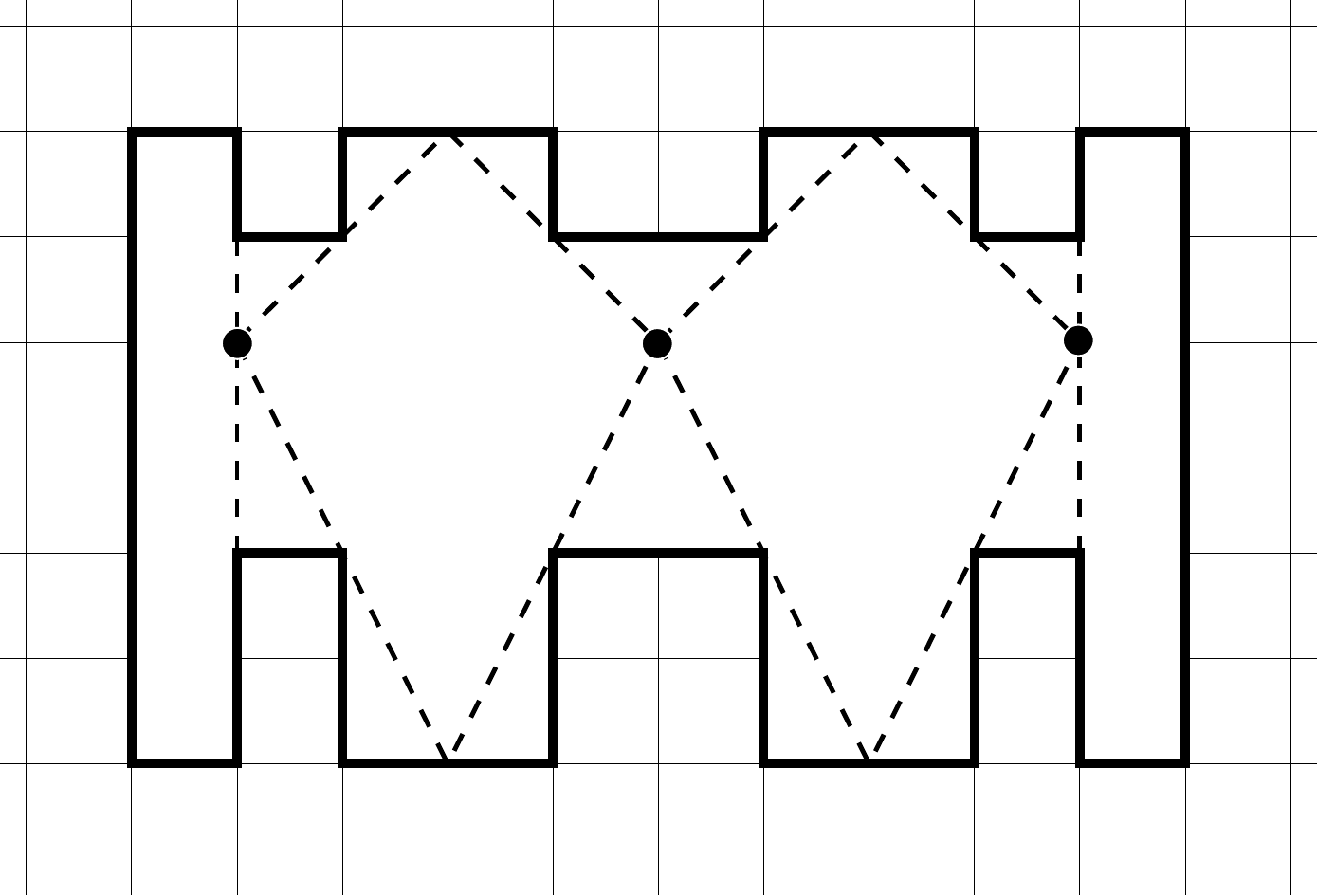}
  \caption{A simple orthogonal polygon possessing only a single
    optimal solution.}
  \label{fig:degenerate}
\end{figure}
The experiments conducted by different groups as well as the results
shown in Section~\ref{sec:experimental_evaluation} indicate that
practically efficient algorithms exist, and a growing number of input
instances can be solved to optimality. This raises the question
whether it can be expected that all instances can be solved, given
sufficient time.

Unfortunately, the answer to this question is ``no''. As a
counterexample, consider the polygon depicted in
Figure~\ref{fig:degenerate}. The three indicated guard positions form
the only optimal solution. There is no variation
allowed---shifting any guard by any $\varepsilon>0$, in an arbitrary
direction, will create a shadow, requiring a fourth guard and thereby
losing optimality.

None of the currently known algorithms  can solve such problems, as no
way  to characterize  these points  is  known. To  see this,  consider
perturbations of  the shown polygon:  It is possible to  slightly move
all vertices in  a way that keeps  the dashed lines intact.  It is not
clear how to  find the shadow alignment points on  the boundary, which
in turn define the optimal guard positions.
\cidC{It should be noted, however,  that it} remains an open  question
whether there are polygons given by rational coordinates  that require
optimal guard positions  \cidC{with irrational coordinates}.\todo{R3.D3}

To summarize, after  forty years of research on \agp,  it is still not
known  whether  there  exist   finite-time  algorithms  for  it.  Even
membership in NP is unclear, as it is not known if guard locations can
be encoded in polynomial size.

%% file: conclusion.tex
\section{Conclusion}
\label{sec:conclusion}

In this paper, we have surveyed recent developments on solving the \agp in a
practically efficient manner. After over thirty years of mostly theoretical work,
several approaches have been proposed and evaluated over the last few years,
resulting in dramatic improvements. The size of instances for which optimal
solutions can be found in reasonable time has improved from tens to thousands of
vertices in just a few years.

While these developments are very promising, experimental findings have
led to new questions about the problem complexity. There are bad
instances that current implementations cannot solve despite small
size, and it is not clear whether exact algorithms for the \agp
can exist, even ones with exponential runtime.